%% file: main.tex
\documentclass[pdftex,twocolumn,epjc3]{svjour3}
\RequirePackage[T1]{fontenc}
\RequirePackage[english]{babel} 

\smartqed
\usepackage{subcaption}
\usepackage{makecell}
\usepackage[utf8]{inputenc}
\usepackage{amsmath}
\usepackage{tabularx}
\usepackage{booktabs}
\usepackage{array}
\RequirePackage{graphicx}
\RequirePackage{mathptmx}      
\RequirePackage{balance}       
\RequirePackage{microtype}     
\usepackage[backend=biber,style=numeric,sorting=none,maxnames=2]{biblatex}
\RequirePackage[colorlinks,citecolor=blue,urlcolor=blue,linkcolor=blue]{hyperref}
\usepackage{float}
\journalname{Eur. Phys. J. C}
\usepackage{adjustbox}
\begin{document}
\sloppy  
\title{Joint LHCb--Belle II Prospects to Constrain New Physics in $B\to D^{(*)}\tau\nu$}

\author{Johannes Albrecht$^1$
        \and
        Florian Bernlochner$^2$
        \and
        Marco Colonna$^1$
        \and
        Lorenz G\"artner$^3$
        \and
        Abhijit Mathad$^4$
        \and
        Biljana Mitreska$^5$
        \and
        Markus Prim$^2$
        \and
        Ilias Tsaklidis$^2$
}

\institute{Fakult\"at Physik, TU Dortmund University, Dortmund, Germany\label{addr1}
          \and
          Physikalisches Institut, Rheinische Friedrich-Wilhelms-Universit\"at Bonn, Bonn, Germany\label{addr2}
          \and
          LMU Munich, Munich, Germany\label{addr5}
          \and
          CERN, Geneva, Switzerland\label{addr3}
          \and
          Department of Physics and Astronomy, University of Manchester, Manchester, United Kingdom\label{addr4}
}

\date{Manuscript to be submitted to Eur.Phys.J.C.}
\maketitle

\newcommand{\btoclnu}{b\to c l\bar{\nu}_{l}}
\newcommand{\btoctaunu}{b\to c\tau\bar{\nu}_{\tau}}
\newcommand{\BtoDorDsttaunu}{\bar{B}\to D^{(*)}\tau\bar{\nu}_{\tau}}
\newcommand{\BtoDorDstellnu}{\bar{B}\to D^{(*)}\ell\bar{\nu}_{\ell}}
\newcommand{\tautomuLHCb}{\tau\to\mu\bar\nu_\mu\nu_\tau}
\newcommand{\tautohadLHCb}{\tau\to 3\pi\nu_\tau}
\newcommand{\tautomuBelle}{\tau\to \ell\bar\nu_{\ell}\nu_{\tau}}
\newcommand{\tautohadBelle}{\tau\to \pi\nu_{\tau}}
\newcommand{\BtoDtaunu}{\bar{B}\to D\tau\bar{\nu}_{\tau}}
\newcommand{\BtoDsttaunu}{\bar{B}\to D^*\tau\bar\nu_{\tau}}
\newcommand{\BtoDststellnu}{\bar{B}\to D^{**}\ell\bar\nu_{\ell}}
\newcommand{\BtoD}{B\!\to\!D}
\newcommand{\BtoDst}{B\!\to\!D^*}
\begin{abstract}
\input{abstract}
\end{abstract}

\input{sec1}

\input{sec2}

\input{sec3}

\input{sec4}

\input{sec5}

\input{sec6}

\section*{Acknowledgements}
\label{sec:acknowledgements}
\input{acknowledgements}

\appendix

\section{Neutrino reconstruction in missing-energy decays}
\label{app:A}

\input{sec7}

\printbibliography

\balance  
\end{document}

%% file: abstract.tex
Semileptonic $\btoctaunu$ decays are powerful probes of non-Standard-Model effects within an effective-field-theory (EFT) framework, but fully exploiting them in current and future data demands combinations that maximise sensitivity while controlling biases from Standard-Model-based modelling and from theory inputs that are shared, and therefore correlated, across analyses in different experiments.
We present a first sensitivity study of a \emph{combined} extraction of Wilson coefficients in $\BtoDorDsttaunu$ decays using LHCb- and Belle~II-like analysis configurations. Detector simulations for signal and backgrounds are typically generated under Standard Model assumptions; if non-SM contributions are present, this can bias the fitted Wilson coefficients. In addition, hadronic inputs such as form-factor parameters of signal and background components are common across analyses, requiring a consistent treatment of fully correlated effects in combinations. To avoid repeating large-scale detector simulation for each EFT hypothesis, we use event-by-event reweighting to map simulated samples to arbitrary combinations of Wilson coefficients. We then compare a simultaneous fit across multiple $\BtoDorDsttaunu$ channels and datasets with a combination based on post-fit averages. Sharing Wilson coefficients and common form-factor parameters in the simultaneous fit reduces model-induced biases and improves sensitivity relative to independent fits, providing a robust and scalable strategy for precision EFT constraints in $\btoctaunu$ transitions using forthcoming LHCb and Belle~II datasets.

%% file: sec1.tex
\section{Introduction}
\label{sec:sec1}

Semileptonic charged-current decays mediated by $\btoclnu$ transitions $l\in\{e,\mu,\tau\}$ constitute a cornerstone of the flavour physics programme, offering stringent tests of the Standard Model (SM). Among these, the modes involving $\tau$ leptons, $\BtoDorDsttaunu$, are particularly incisive: potential non-SM contributions can interfere with the SM amplitude, leaving characteristic imprints on both decay rates and kinematic distributions. Measurements of the lepton-flavour-universality ratios, $R(D^{(*)})=\mathcal{B}(\BtoDorDsttaunu)/\mathcal{B}(\BtoDorDstellnu)$ with $\ell\in\{e,\mu\}$, have been performed by BaBar~\cite{BaBar:2012obs,BaBar:2013mob}, Belle~\cite{Belle:2015qfa,Belle:2016dyj,Belle:2016ure,Belle:2019rba}, Belle~II~\cite{Belle-II:2024ami} and LHCb~\cite{LHCb:2017rln,LHCb:2023zxo,LHCb:2015gmp,LHCb:2017smo,LHCb:2023uiv}. The current averages compiled by HFLAV~\cite{HFLAV:2025CKM} indicate a combined deviation of approximately $3.8\,\sigma$ from SM predictions. This persistent anomaly has established $\btoctaunu$ transitions as a prime target for precision studies and interpretations within the Weak Effective Theory (WET), where New Physics (NP) effects are encoded in four-fermion operators and their associated Wilson coefficients.

As experimental precision improves, the methodology used to extract and combine these measurements must evolve to match the complexity of global WET fits. Contemporary global analyses, facilitated by public software packages~\cite{vanDyk:2022eos,Straub:2018flavio,Aebischer:2018wilson}, combine $R(D^{(*)})$ with polarisation observables and complementary decay channels to constrain Wilson coefficients~\cite{Murgui:2019czp,Fedele:2022iib,Iguro:2024hyk}. However, the validity of these combinations relies on the assumption that the input experimental likelihoods remain accurate when scanning far beyond the SM point. This assumption faces two significant challenges.

The first challenge arises when combining multiple datasets. Theory inputs, particularly the hadronic form-factor parameters, are common across different analyses and experiments. Treating these parameters via post-fit averages or partially correlated nuisance parameters can obscure the propagation of information between datasets, leading to inconsistent constraints. A robust combination requires simultaneous fits where Wilson coefficients and shared hadronic parameters are explicitly correlated across all channels.

The second challenge is model dependence in signal extraction. Measurements of $\btoctaunu$ typically rely on template fits to reconstructed distributions, such as the missing reconstructed mass of the event, $m^2_{\rm miss}$, where signal and background shapes are derived from Monte Carlo (MC) simulations assuming SM couplings. If NP modifies the underlying kinematic distributions, the experimental acceptance, resolution effects and template shapes change coherently. Consequently, reinterpreting SM-derived results in terms of Wilson coefficients can introduce bias. This issue necessitates model-agnostic likelihood approaches, a strategy recently exemplified by the Belle~II analysis of the rare decay $B^{+}\to K^{+}\nu\bar{\nu}$~\cite{BelleII:2025mal,G_rtner_2024}.

To address the first challenge, we perform a first sensitivity study of a \emph{combined} extraction of WET Wilson coefficients using LHCb- and Belle~II-like analysis configurations. We compare a simultaneous multi-channel, multi-dataset fit, sharing Wilson coefficients and common form-factor parameters at the likelihood level, to a combination strategy based on post-fit averages. This allows us to quantify the impact of consistently treating shared theory parameters on model-induced bias and on the sensitivity of combined WET constraints. This approach can be readily expanded to include other common parameters, such as shared systematic uncertainties. 

To address the second challenge, we introduce \textsc{ReDist}~\cite{Scipy2025_Redist}, a framework that embeds simulation reweighting directly into the likelihood minimization (via an interface to \textsc{pyhf}~\cite{pyhf}). This ensures that theory-dependent variations -- such as New Physics modifications to kinematic distributions -- are consistently propagated through detector acceptance, resolution effects and template shapes. While analogous functionality exists within RooFit~\cite{Verkerke:2003ir} workflows (e.g., the \textsc{RooHammerModel}~\cite{GarciaPardinas:2022rhm}), our architecture offers distinct advantages: it facilitates the seamless interchange of theory backends -- such as \textsc{HAMMER}~\cite{Bernlochner_2020} or \textsc{EOS}~\cite{vanDyk:2022eos} -- for robust systematic cross-checks, and provides a portable, reproducible statistical model suitable for modern global combinations.

The structure of this paper is as follows. Section~\ref{sec:sec2} introduces the WET parameterisation and the analysis strategy for $B\to D^{(*)}\tau\nu$. Section~\ref{sec:sec3} describes the \textsc{ReDist} likelihood construction and its interface to \textsc{HAMMER}. Section~\ref{sec:sec4} presents the simulated datasets and experimental configurations. Section~\ref{sec:sec5} reports the results of the combination studies, and Section~\ref{sec:sec6} summarizes the implications and provides an outlook.

%% file: sec2.tex
\section{Theoretical framework and WET parameterisation}
\label{sec:sec2}

Following the motivation outlined in the Introduction, we interpret $b\to c\tau\bar{\nu}_\tau$ transitions in the Weak Effective Theory (WET) framework~\cite{PhysRevD.96.036012,Tanaka:2013ive,Sakaki:2013bfa}. Possible New Physics (NP) effects are parameterised as deviations from the SM only in the $\tau$ sector, while $b\to c\ell\bar{\nu}_\ell$ decays with $\ell=e,\mu$ are treated as SM-like. Assuming no right-handed neutrinos\footnote{Extending the example to BSM coupling with right-handed neutrinos is straightforward.}, the effective Hamiltonian at $\mu\simeq m_b$ is
\begin{equation}
\centering
\mathcal{H}_{\rm eff}
=
\frac{4G_F}{\sqrt{2}}\,V_{cb}
\Big[
\mathcal{O}_{V_L}
+
\vec C \cdot \vec{\mathcal{O}}
\Big]
+{\rm h.c.},
\qquad
\vec C_{\rm SM}=\vec 0,
\label{eq:heff}
\end{equation}
where $\vec C\equiv(C_{V_L},C_{V_R},C_{S_L},C_{S_R},C_T)$ denotes Wilson-coefficient shifts with respect to the SM, and
$\vec{\mathcal{O}}\equiv(\mathcal{O}_{V_L},\mathcal{O}_{V_R},\mathcal{O}_{S_L},\mathcal{O}_{S_R},\mathcal{O}_T)$.
The operator basis is
\begin{equation}
\centering
\begin{aligned}
\mathcal{O}_{V_L} &= (\bar c\gamma^\mu P_L b)(\bar\tau\gamma_\mu P_L\nu_\tau), \qquad
&\mathcal{O}_{V_R} &= (\bar c\gamma^\mu P_R b)(\bar\tau\gamma_\mu P_L\nu_\tau), \\
\mathcal{O}_{S_L} &= (\bar c P_L b)(\bar\tau P_L\nu_\tau), \qquad
&\mathcal{O}_{S_R} &= (\bar c P_R b)(\bar\tau P_L\nu_\tau), \\
\mathcal{O}_{T}   &= (\bar c\sigma^{\mu\nu} P_L b)(\bar\tau\sigma_{\mu\nu} P_L\nu_\tau), &&
\end{aligned}
\label{eq:wet_ops}
\end{equation}
with $P_{L,R}=(1\mp\gamma_5)/2$ and $\sigma^{\mu\nu}=\tfrac{i}{2}[\gamma^\mu,\gamma^\nu]$. In this parameterisation, non-zero Wilson coefficients modify both the overall rate and the differential decay structure through interference with the SM amplitude, providing sensitivity beyond inclusive observables such as $R(D^{(*)})$.

The short-distance physics is universal: the same Wilson-coefficient vector $\vec C$ governs every $b\to c\tau\bar{\nu}_\tau$ transition. By contrast, QCD effects enter through hadronic matrix elements, which are mode dependent (though related across channels by heavy-quark symmetries). Whenever the same decay modes contribute to multiple analyses -- either as signal or as background -- these hadronic inputs must be treated consistently and therefore shared across fits. This is particularly relevant for $B\to D^{(*)}$ transitions, whose hadronic matrix elements are encoded in form factors. We collect the corresponding form-factor normalisations and shape parameters in a vector $\vec\alpha$, using for instance CLN- or dispersive $z$-expansion parameterisations~\cite{Caprini:1997mu,Boyd:1994tt,Boyd:1995cf,Boyd:1995sq,Boyd:1997kz}, as well as HQET-improved treatments retaining subleading Isgur--Wise functions (often referred to as the BLPR(XP) approach)~\cite{Bernlochner:2017jka,Bernlochner:2022hqe}. Where needed, the normalisations (and, increasingly, shape information) can be anchored by lattice-QCD inputs~\cite{Bailey:2014tva,Bailey:2015dka}.

\begin{figure}
    \centering
    \includegraphics[width=0.8\linewidth]{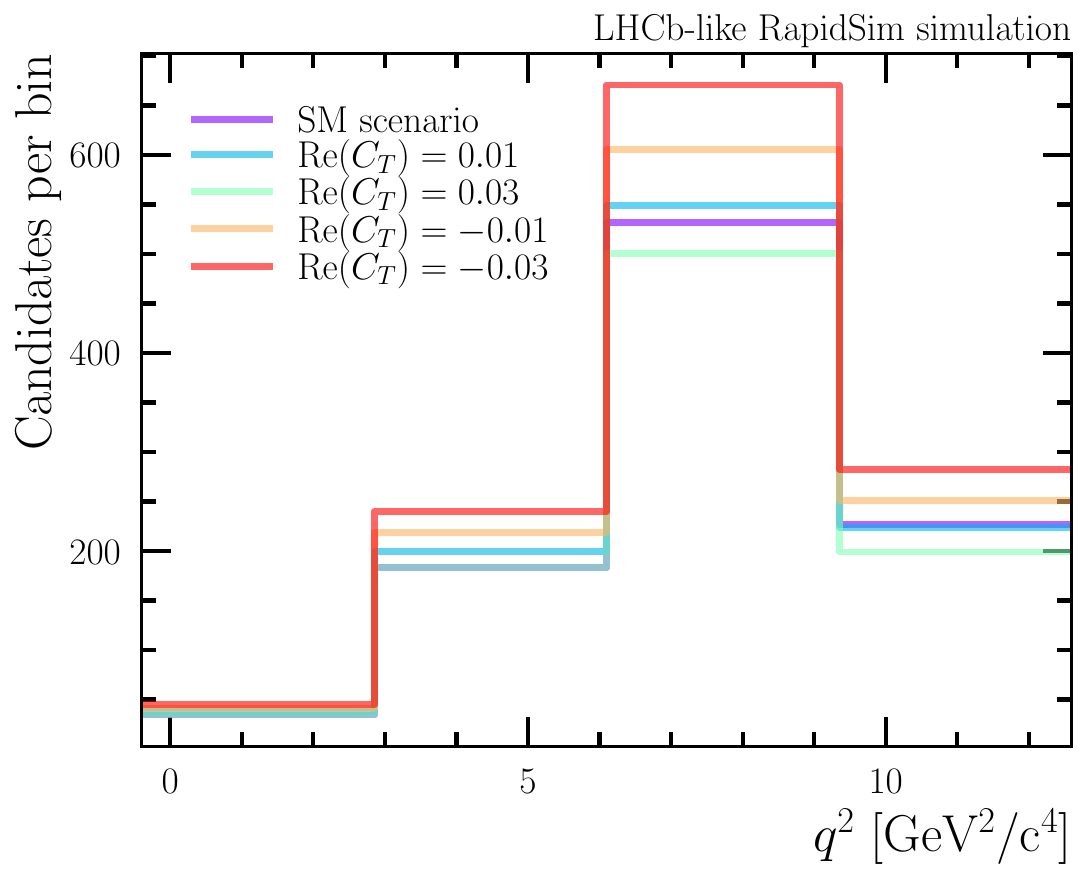}
    \hfill
    \includegraphics[width=0.8\linewidth]{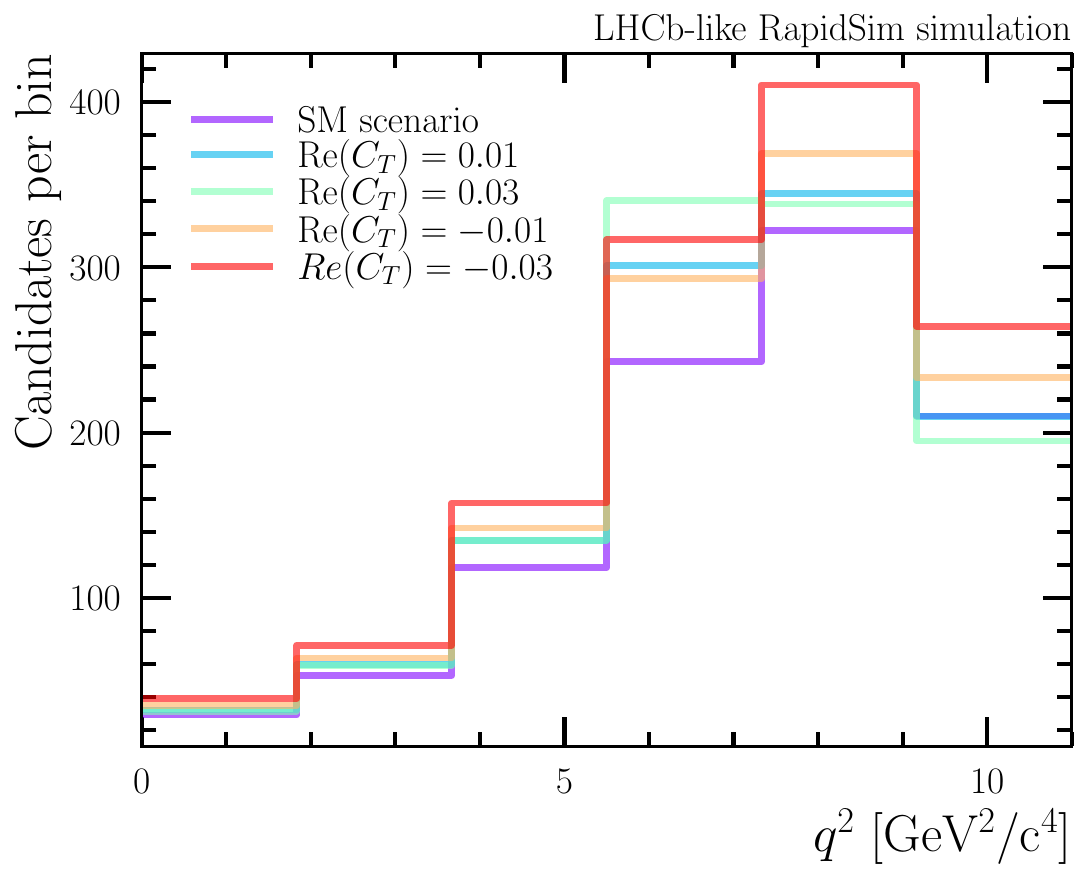}
    \caption{Distributions of the reconstructed squared momentum transfer $q^2$ for the SM (purple) and illustrative variations of $\mathrm{Re}(C_T)$ (orange). Shown are LHCb-like simulated decays for $B\to D^*\tau(\to\mu\nu\nu)\nu$ (top) and $B\to D^*\tau(\to3\pi\nu)\nu$ (bottom).
    }
    \label{fig:Hammer_weight}
\end{figure}

Figure~\ref{fig:Hammer_weight} illustrates a representative effect on a key kinematic variable, the squared momentum transfer $q^2$, for LHCb-like simulated $B\to D^*\tau\bar{\nu}_\tau$ decays. Variations of $\mathrm{Re}(C_T)$ distort the $q^2$ spectrum, exemplifying the coherent shape changes that motivate the likelihood-level strategy described next~\cite{Bernlochner:2020qhk}.

%% file: sec3.tex
\section{Likelihood construction with \textsc{ReDist} and \textsc{HAMMER}}
\label{sec:sec3}

To address the model-dependence issue highlighted in the Introduction, we construct a likelihood in which the reconstructed signal model depends explicitly on the shared parameter set $\theta=(\vec C,\vec\alpha)$, avoiding the need to regenerate detector simulation for each theory hypothesis. We employ a \textsc{HistFactory}-style binned likelihood~\cite{Cranmer:1456844} implemented in \textsc{pyhf}~\cite{pyhf}, and evaluate theory variations using event-by-event matrix-element reweighting with \textsc{HAMMER}~\cite{Bernlochner_2020}. The interface is provided by \textsc{ReDist}~\cite{G_rtner_2024,Scipy2025_Redist}, which promotes theory-driven template deformations to first-class objects in the statistical model, enabling coherent multi-channel combinations where $\vec C$ and $\vec\alpha$ are floated simultaneously.

Starting from a nominal fully simulated sample generated at $\theta_0\equiv(\vec C_0,\vec\alpha_0)$, each event $e$ with truth-level phase-space coordinates $k_e$ is assigned a theory weight
\begin{equation}
\centering
w_e(\theta)
=
\frac{{\rm d}\Gamma/{\rm d}\Omega\,(\theta; k_e)}
     {{\rm d}\Gamma/{\rm d}\Omega\,(\theta_0; k_e)}\,,
\label{eq:we}
\end{equation}
where $\Omega$ denotes the set of phase-space variables defining the differential rate. Because the weights are applied to fully simulated events, changes in the decay dynamics induced by $\theta$ propagate through the detector response and reconstruction, producing consistent modifications of acceptance- and resolution-smeared templates.

The expected signal yield $\nu_b$ in a reconstructed analysis bin $b$ is then computed as
\begin{equation}
\centering
\nu_b(\theta)
=
\sum_{e\in b} s_e\, w_e(\theta),
\label{eq:nub}
\end{equation}
where $s_e$ denotes simulation-to-data scaling factors (including normalisation and, if applicable, efficiency weights). For implementation in \textsc{pyhf}, it is convenient to factorise the prediction into a nominal template and a morphing function,
\begin{equation}
\centering
\nu_b(\theta)
=
\nu_b(\theta_0)\, r_b(\theta),
\qquad
r_b(\theta)
=
\frac{\sum_{e\in b} s_e\, w_e(\theta)}
     {\sum_{e\in b} s_e}\,,
\label{eq:rb}
\end{equation}
so that all dependence on $\theta$ enters through $r_b(\theta)$, while the baseline experimental modelling remains encoded in $\nu_b(\theta_0)$.

The \textsc{ReDist} design is backend-agnostic, allowing alternative theory engines (e.g.\ \textsc{EOS}~\cite{vanDyk:2022eos}) to be substituted for validation without changing the statistical model. This enables scalable, high-dimensional scans of $\theta$ within a single, portable likelihood description, suited to the combined sensitivity studies presented in the following sections.

%% file: sec4.tex
\section{Analysis configurations and template models}
\label{sec:sec4}

We construct simplified LHCb- and Belle~II-like template models to evaluate the sensitivity of combined likelihood fits to the WET parameter set $\theta=(\vec C,\vec\alpha)$. The models are implemented through the \textsc{ReDist}--\textsc{HAMMER} interface as \textsc{HistFactory}/\textsc{pyhf} likelihoods, both per analysis and in combination. Unless stated otherwise, \textit{Asimov} datasets are generated at the nominal point $\theta_0$, and the fits float the Wilson coefficients and the shared form-factor parameters, so that theory hypotheses coherently deform the reconstructed templates.

The analysis channels, included signal/background components, and reconstructed fit observables are summarised in Table~\ref{tab:channels_summary}. The background composition is intentionally streamlined relative to the full experimental measurements: sub-dominant sources that are not needed for the sensitivity studies, or that are largely shape-degenerate in the observables used here, are omitted; where appropriate, background \emph{classes} are represented by a single proxy decay mode.

\begin{table*}[t]
\centering
\footnotesize
\setlength{\tabcolsep}{5pt}
\renewcommand{\arraystretch}{1.18}
\caption{Summary of the LHCb- and Belle~II-like template models used in this work: analysis channels, included physics components, and reconstructed observables entering the binned likelihood fits. Charge-conjugate modes are implied throughout.}
\label{tab:channels_summary}
\begin{tabularx}{\textwidth}{
>{\raggedright\arraybackslash}p{1.8cm}
>{\raggedright\arraybackslash}p{4.2cm}
>{\raggedright\arraybackslash}X
>{\raggedright\arraybackslash}p{4.4cm}
}
\toprule
\textbf{Experiment} &
\textbf{Signal region / $\tau$ channel} &
\textbf{Included components} &
\textbf{Fit observables}
\\
\midrule

LHCb-like &
$R(D)$, $\tautomuLHCb$ &
\textbf{Signal:} $B^0\to D^-\tau^+\nu_\tau$ \newline
\textbf{Normalisation:} $B^0\to D^-\mu^+\nu_\mu$ \newline
\textbf{Feed-down:} $B^0\to D^{**-}(2420,2430,2460)\mu^+\nu_\mu$; \;
$B^0\to D^{*-}\mu^+\nu_\mu$ (down-feed, unreconstructed $D^*$)
&
$m^2_{\rm miss}$,\; $q^2$,\; $E^{\rm FS}_{\rm lep}$
\\
\midrule

LHCb-like &
$R(D^{*})$, $\tautomuLHCb$ &
\textbf{Signal:} $B^0\to D^{*-}\tau^+\nu_\tau$ \newline
\textbf{Normalisation:} $B^0\to D^{*-}\mu^+\nu_\mu$ \newline
\textbf{Feed-down:} $B^0\to D^{**-}(2420,2430,2460)\mu^+\nu_\mu$ \newline
\textbf{Double-charm (proxy):} $B^0\to D^{*-}D_s^{*+}$
&
$m^2_{\rm miss}$,\; $q^2$,\; $E^{\rm FS}_{\rm lep}$
\\
\midrule

LHCb-like &
$R(D^{*})$, $\tautohadLHCb$ &
\textbf{Signal:} $B^0\to D^{*-}\tau^+\nu_\tau$ \newline
\textbf{Feed-down:} $B^0\to D^{**-}(2420,2430,2460)\tau^+\nu_\tau$ \newline
\textbf{Double-charm (proxy):} $B^0\to D^{*-}D_s^{*+}$
&
$q^2$,\; $m_{\pi\pi}^{\rm MIN}$,\; $m_{\pi\pi}^{\rm MAX}$
\\
\midrule

Belle~II-like &
$R(D)$, $\tautomuBelle$ ($\ell=e,\mu$) &
\textbf{Signal:} $B^{0(+)}\to D^{-(0)}\tau^+\nu_\tau$ \newline
\textbf{Normalisation:} $B^{0(+)}\to D^{-(0)}\ell^+\nu_\ell$ \newline
\textbf{Down-feed:} $B^{0(+)}\to D^{*-(*)}\ell^+\nu_\ell$ (mis-reconstructed as $D$) \newline
\textbf{Hadronic (proxy):} $B^{0(+)}\to D^{-(0)}\pi^-\pi^+\pi^+\pi^0$
&
$m^2_{\rm miss}$,\; $E_{\rm extra}$
\\
\midrule

Belle~II-like &
$R(D^{*})$, $\tautomuBelle$ ($\ell=e,\mu$) &
\textbf{Signal:} $B^{0(+)}\to D^{*-(*)}\tau^+\nu_\tau$ \newline
\textbf{Normalisation:} $B^{0(+)}\to D^{*-(*)}\ell^+\nu_\ell$ \newline
\textbf{$D^{**}$ feed-down (proxy):} $B^{0(+)}\to D^{**-(0)}(2420)\,\ell^+\nu_\ell$ \newline
\textbf{Hadronic (proxy):} $B^{0(+)}\to D^{*-(*)}\pi^-\pi^+\pi^+\pi^0$
&
$m^2_{\rm miss}$,\; $E_{\rm extra}$
\\
\midrule

Belle~II-like &
$R(D^{*})$, $\tautohadBelle$ (one-prong hadronic) &
\textbf{Signal:} $B^{0(+)}\to D^{*-(*)}\tau^+\nu_\tau$ \newline
\textbf{$D^{**}$ feed-down (proxy):} $B^{0(+)}\to D^{**-(0)}(2420)\,\ell^+\nu_\ell$ \newline
\textbf{Hadronic (proxy):} $B^{0}\to D^{*-}\pi^+\pi^+\pi^-\pi^0$
&
$m^2_{\rm miss}$,\; $E_{\rm extra}$
\\
\bottomrule
\end{tabularx}

\vspace{2pt}
\begin{minipage}{0.98\textwidth}
\footnotesize
\textit{Notes:} Signal and normalisation $B\to D^{(*)}(\tau,\ell)\nu$ modes are processed with \textsc{HAMMER} and reweighted to the \textit{BLPRXP} form-factor hypothesis~\cite{Bernlochner_2022}. For the LHCb-like configuration, $B\to D^{**}\ell\nu$ feed-down modes use the \textit{BLR} parameterisation~\cite{Bernlochner:2017jxt} with Gaussian constraints on the corresponding form-factor parameters. At Belle~II, the $\tau\to\pi\nu$ category is used as a proxy for an inclusive one-prong hadronic sample; since $\tau\to\rho\nu$ reweighting is not yet available in our setup, the effective $\tau\to\pi\nu$ signal statistics are increased to approximately account for the missing contribution.
\end{minipage}
\end{table*}

\subsection{LHCb-like configuration}
\label{sec:sec4:lhcb}

LHCb-like samples are generated with \textsc{RapidSim}~\cite{Cowan_2017}, which emulates acceptance and resolution effects via Gaussian smearing of final-state kinematics. Because the final state contains neutrinos, the signal-side kinematics are reconstructed using LHCb-inspired approximations: a $B$-rest-frame boost approximation for $\tau\to\mu\bar\nu_\mu\nu_\tau$, and a two-fold quadratic ambiguity method for $\tau\to3\pi\nu_\tau$; details are given in Appendix~\ref{app:A}.

The $\tau$-muonic configurations emulate the dominant ingredients of the public LHCb $R(D)$ and $R(D^{*})$ analyses~\cite{LHCb:2023zxo} by fitting the $\tau$ signal modes simultaneously with the corresponding $b\to c\mu\bar\nu_\mu$ normalisation modes. In this setup the normalisation channel yield is chosen to be $\mathcal{O}(20)$ times larger than the signal, reflecting the experimental hierarchy and ensuring that the shared form-factor parameters are tightly constrained by data. The $\tau$-hadronic configuration emulates the LHCb $R(D^{*})$ analysis with $\tautohadLHCb$~\cite{LHCb:2017rln}; in isolation this mode provides weaker constraints on the $B\to D^{*}$ form factors, since the experimental normalisation is derived from a topology that is not included here. In the combined fits of Section~\ref{sec:sec5}, the Wilson coefficients and the signal form-factor parameters are shared across $\tau$-muonic and $\tau$-hadronic samples, allowing the normalisation-driven constraints to propagate consistently.

\begin{figure*}[!t]
  \centering
  \setlength{\tabcolsep}{5pt}
  \renewcommand{\arraystretch}{1.0}
  \vspace{-1mm}
  \begin{tabular}{c c c c}
    & \textbf{$q^2$} &
      \textbf{$E^{\rm FS}_{\rm lep}$ / $m_{\pi\pi}^{\rm MIN}$} &
      \textbf{$m^2_{\rm miss}$ / $m_{\pi\pi}^{\rm MAX}$} \\
    \multicolumn{4}{c}{\rule{0pt}{2.2ex}} \\[-0.5ex]

    \rotatebox{90}{\hspace{3.0em}\textbf{$R(D^{*})$, $\tau\!\to\!\mu\nu\nu$}} &
    \includegraphics[width=0.29\textwidth]{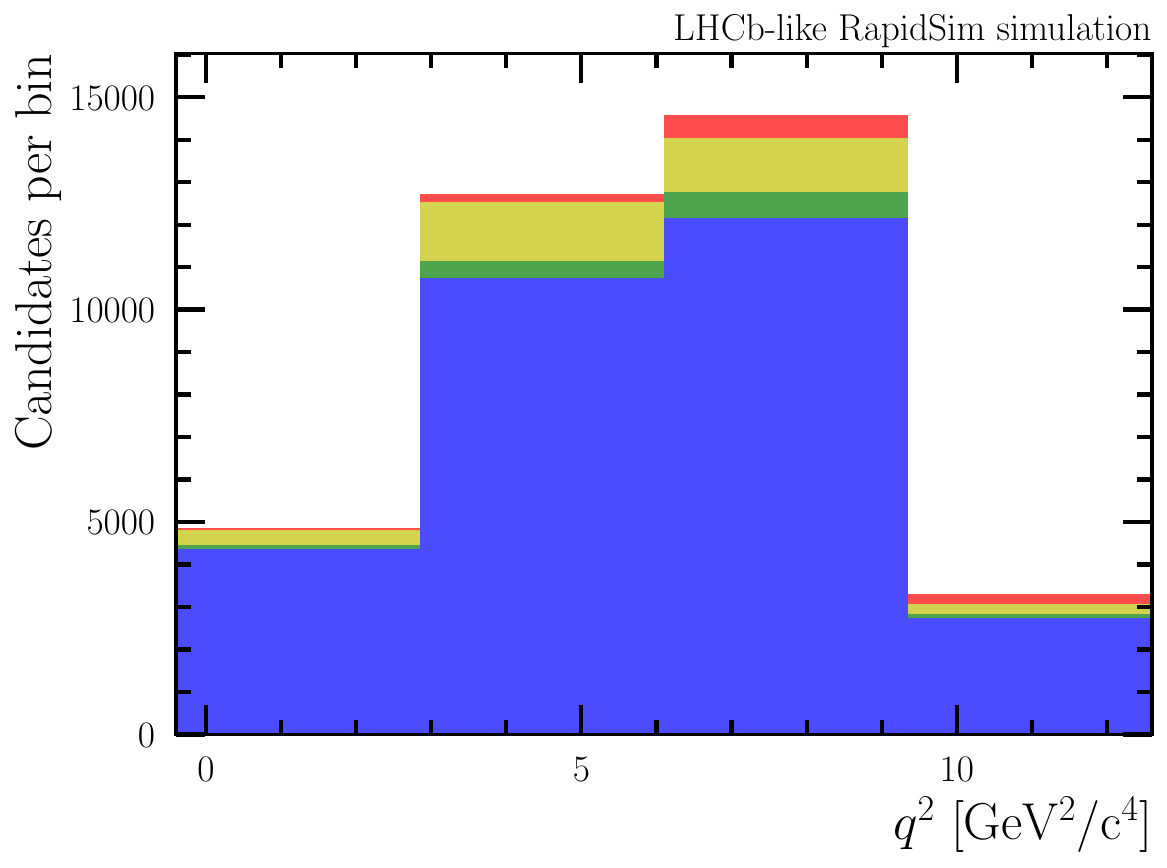} &
    \includegraphics[width=0.29\textwidth]{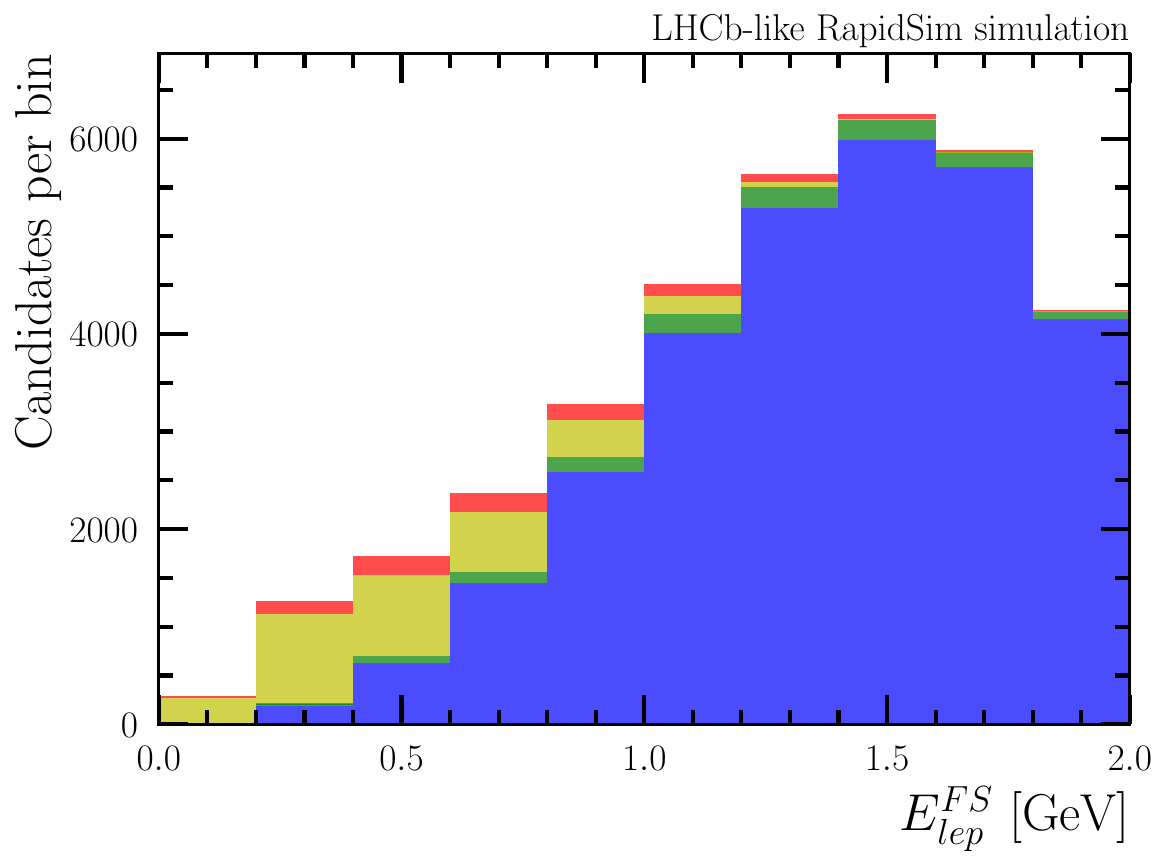} &
    \includegraphics[width=0.29\textwidth]{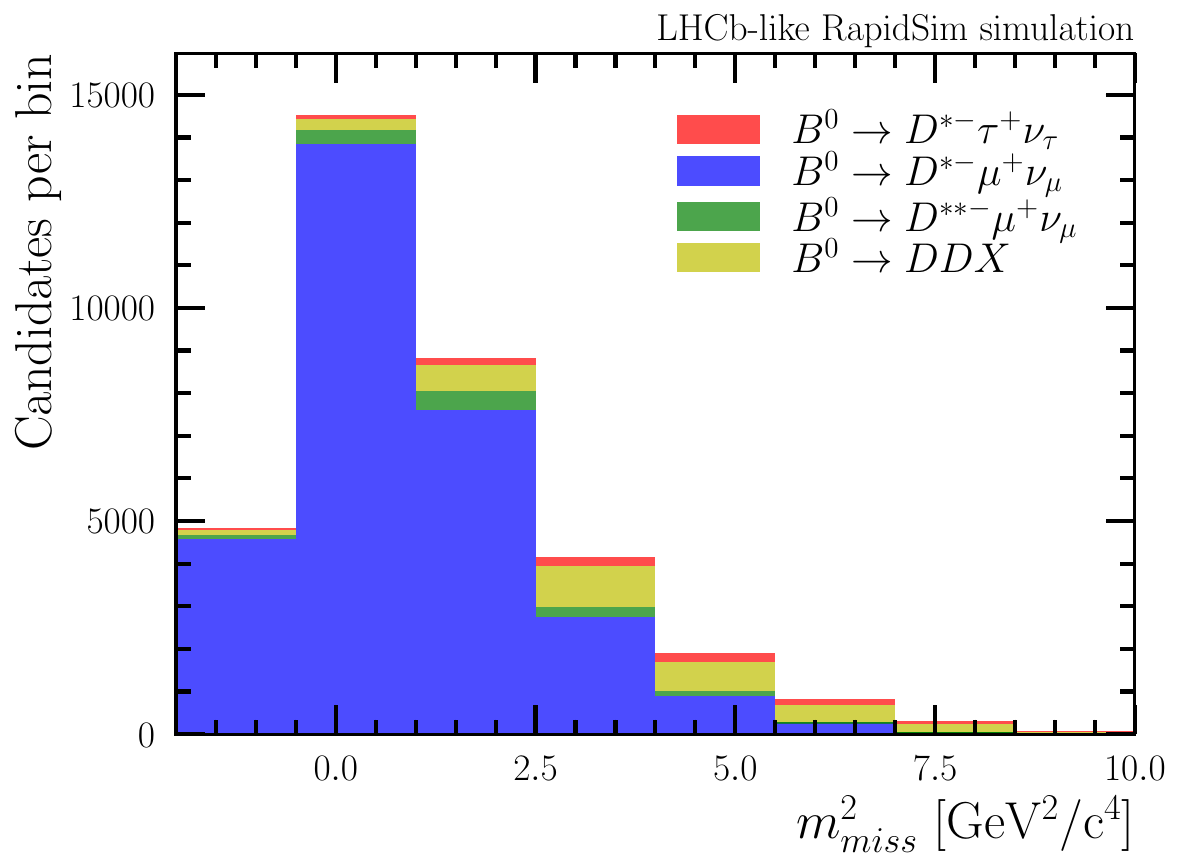} \\[0.6ex]

    \rotatebox{90}{\hspace{3.0em}\textbf{$R(D)$, $\tau\!\to\!\mu\nu\nu$}} &
    \includegraphics[width=0.29\textwidth]{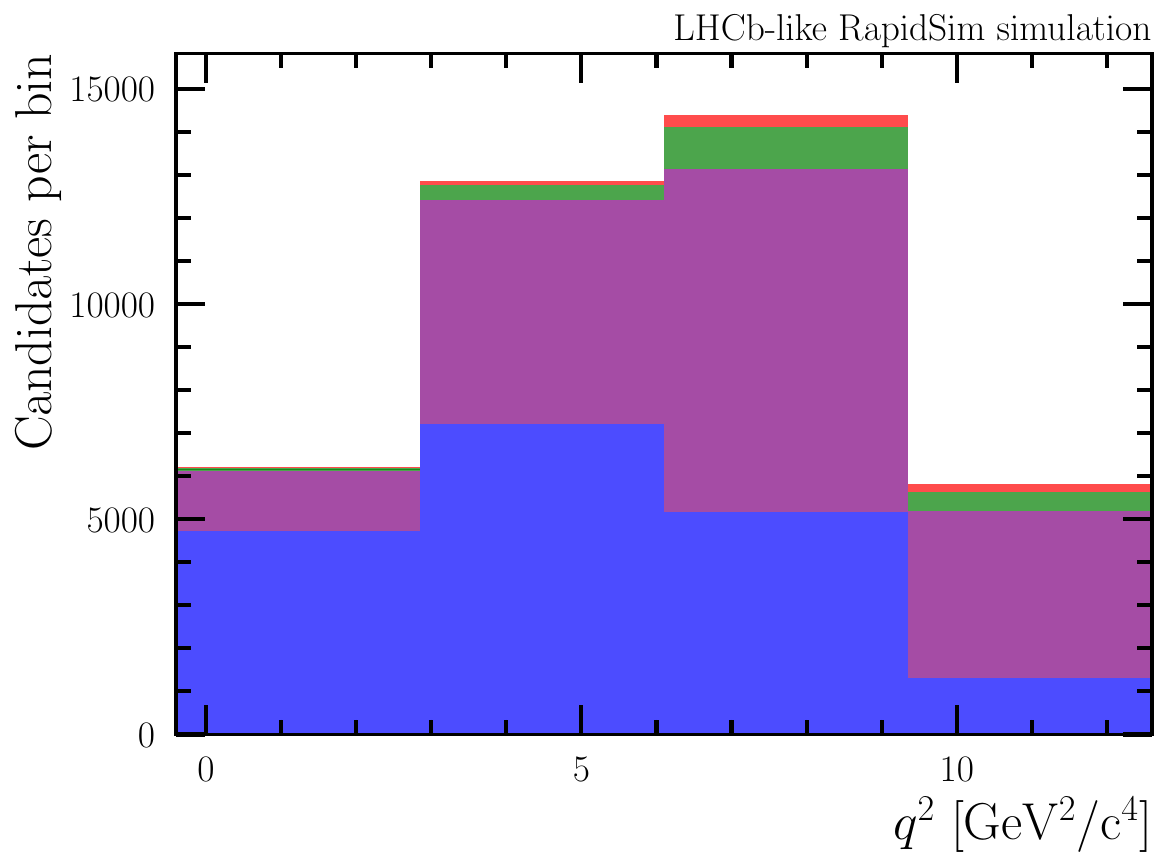} &
    \includegraphics[width=0.29\textwidth]{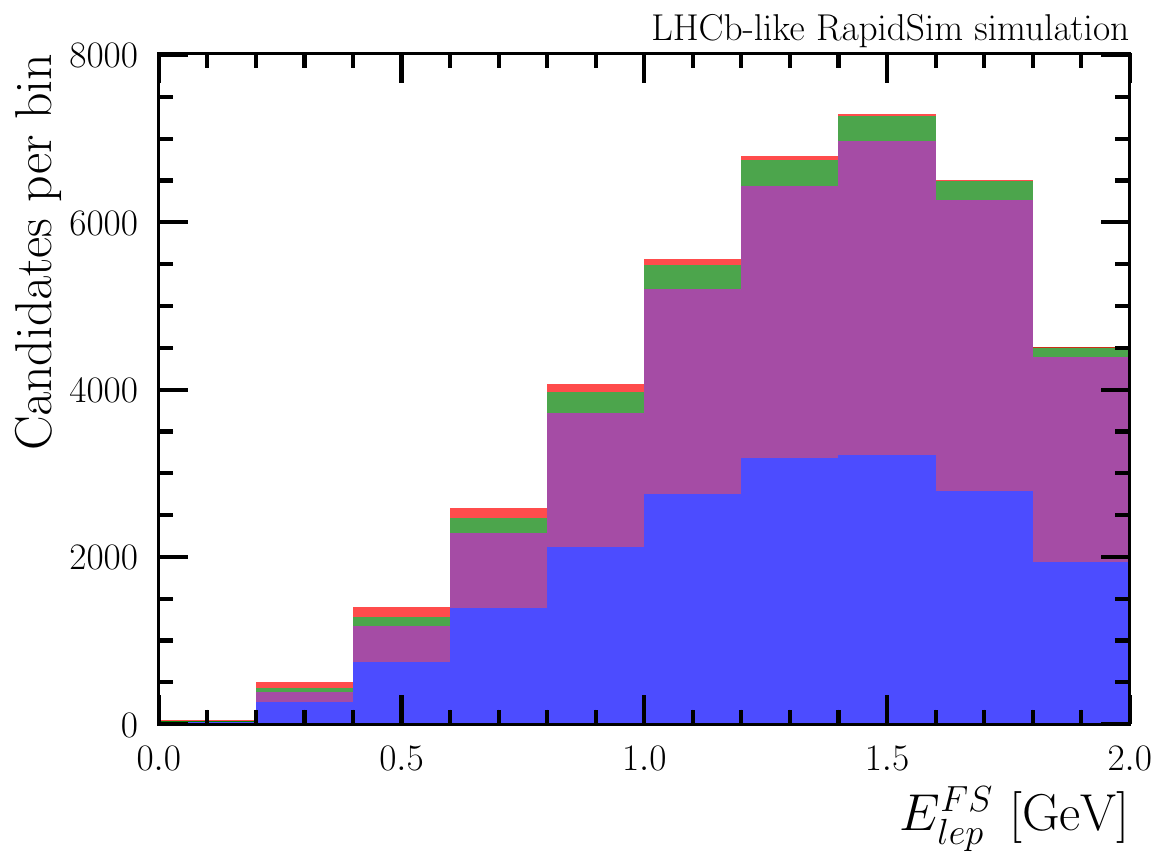} &
    \includegraphics[width=0.29\textwidth]{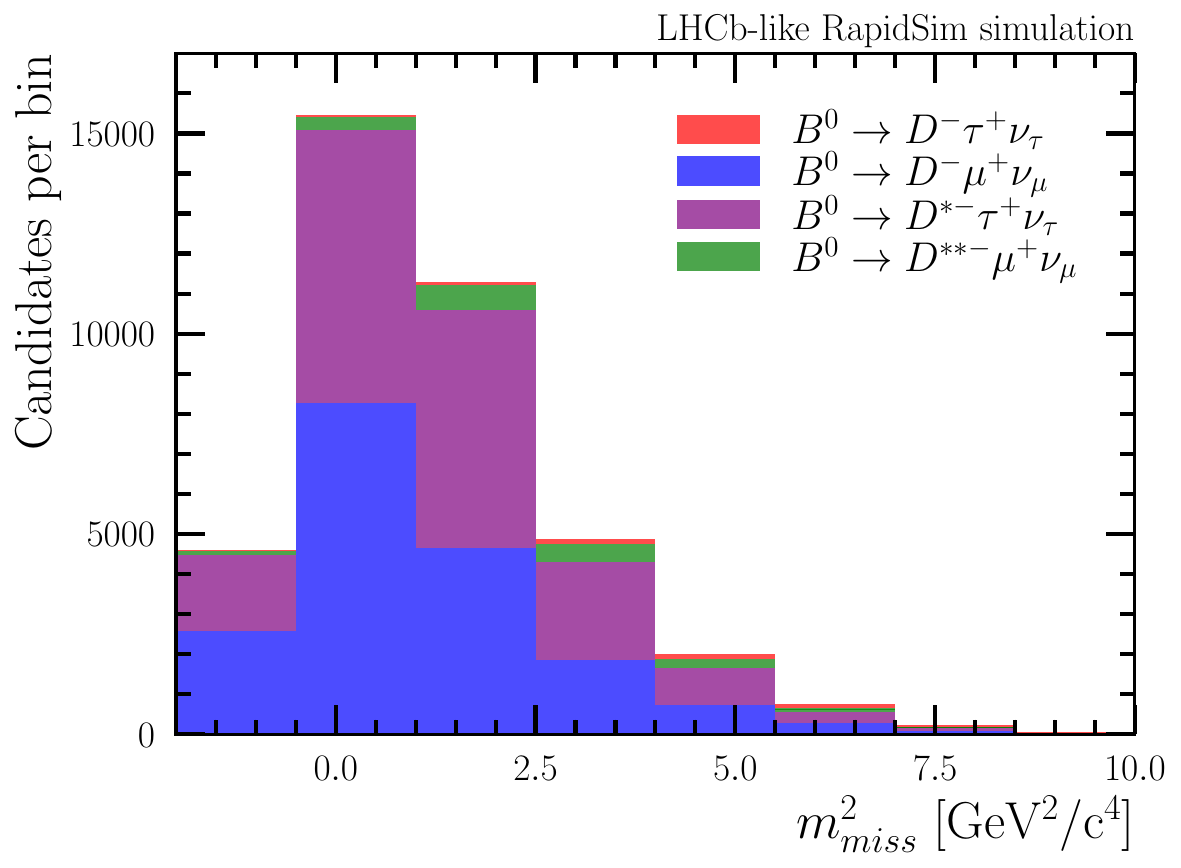} \\[0.6ex]

    \rotatebox{90}{\hspace{3.0em}\textbf{$R(D^{*})$, $\tau\!\to\!3\pi\nu$}} &
    \includegraphics[width=0.29\textwidth]{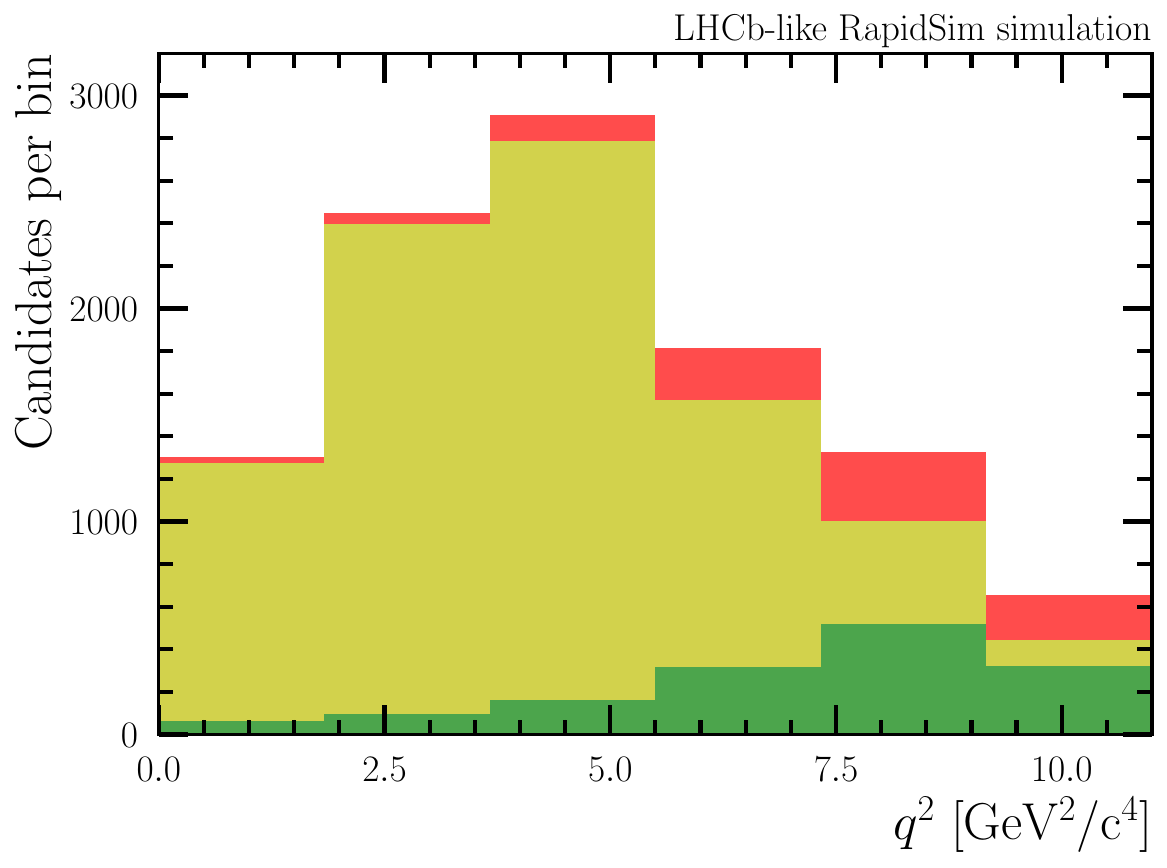} &
    \includegraphics[width=0.29\textwidth]{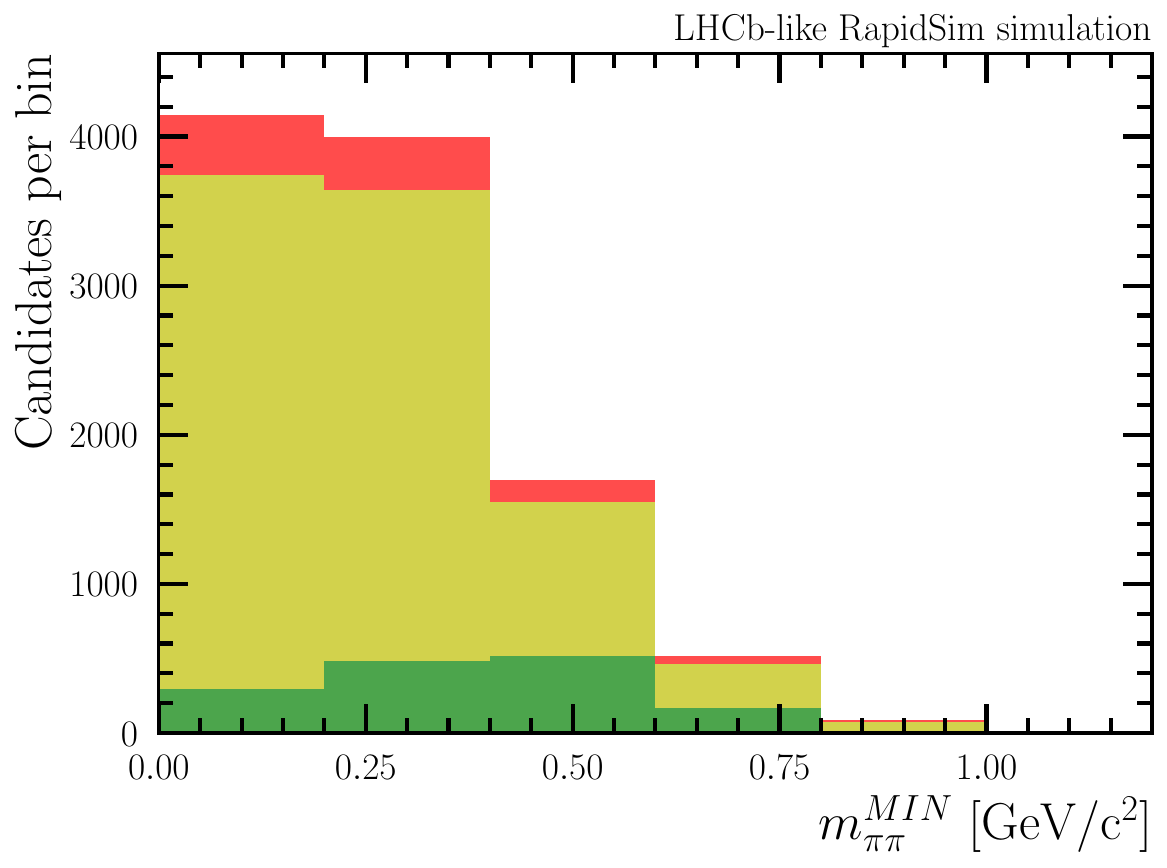} &
    \includegraphics[width=0.29\textwidth]{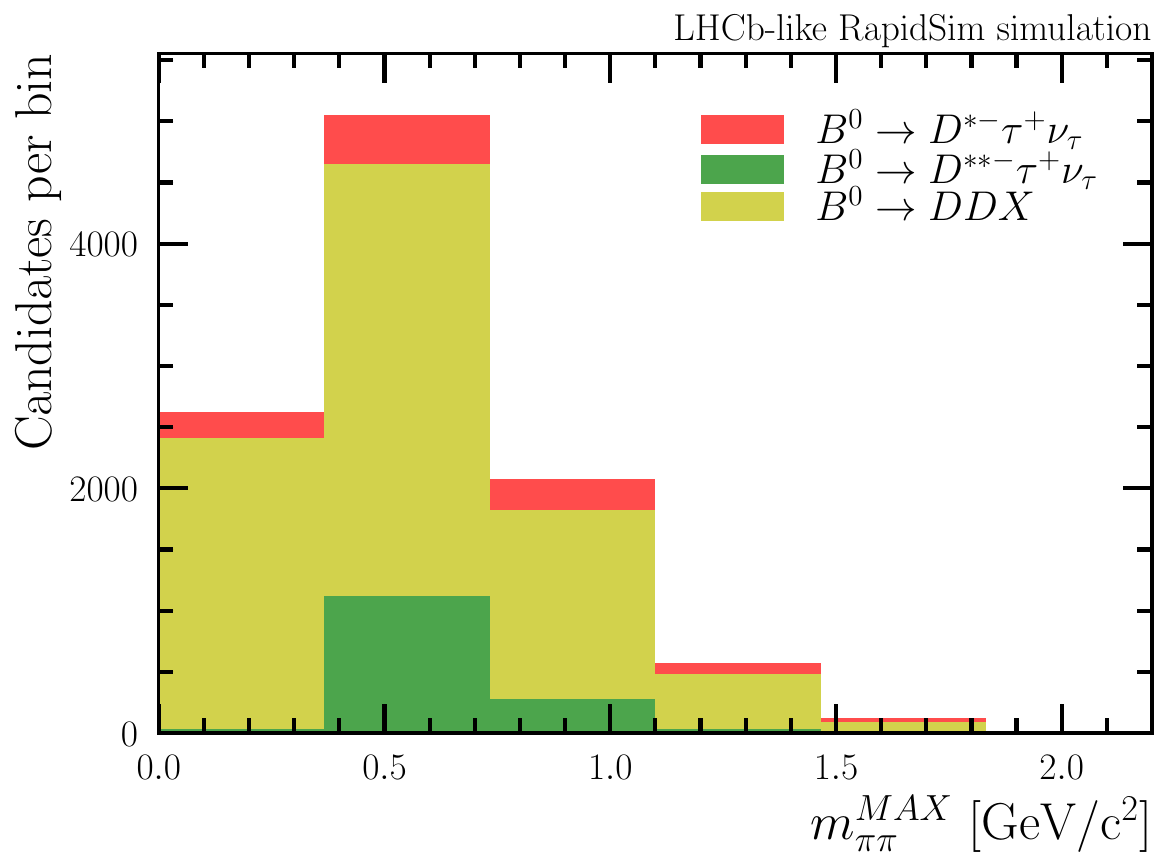} \\
  \end{tabular}
  
  \caption{LHCb-like template projections arranged by channel (rows) and observable (columns). For $\tau\to\mu\bar\nu_\mu\nu_\tau$, the columns show the reconstructed $(q^2,\,E^{\rm FS}_{\rm lep},\,m^2_{\rm miss})$ for the $R(D^{*})$ and $R(D)$ signal regions. For $\tau\to3\pi\nu_\tau$, the three panels correspond to $(q^2,\,m_{\pi\pi}^{\rm MIN},\,m_{\pi\pi}^{\rm MAX})$. Signal is shown in red; other colours indicate background components.}
  \label{fig:LHCbAllPanels}
\end{figure*}

For the dominant feed-down from $\BtoDststellnu$ we use the \textit{BLR} parameterisation~\cite{Bernlochner:2017jxt} with Gaussian constraints on its form-factor parameters. For the remaining backgrounds we retain only the leading double-charm class, represented by a single proxy mode $B^0\to D^{*-}D_s^{*+}$; smaller contributions present in the full analyses (combinatorial, misidentification, and minor feed-down components) are neglected. The resulting LHCb-like template projections are shown in Fig.~\ref{fig:LHCbAllPanels}; the columns are ordered as $(q^2,\;E^{\rm FS}_{\rm lep}\!/m_{\pi\pi}^{\rm MIN},\;m^2_{\rm miss}\!/m_{\pi\pi}^{\rm MAX})$ to provide a consistent layout across the muonic and hadronic $\tau$ categories.

\subsection{Belle~II-like configuration}
\label{sec:sec4:belle2}

The Belle~II-like configuration exploits the clean $e^+e^-$ environment and the kinematic constraints provided by hadronic $B$-tagging~\cite{fei}. We implement a simplified tagging scenario in which the tag-side $B$ meson decays as $B^0\to D^-\pi^+$ or $B^+\to\bar D^0\pi^+$ and is assumed to be reconstructed perfectly. Signal-side and background decays are generated with \textsc{EvtGen}~\cite{Lange:2001uf}, following the strategies used in Belle~\cite{Belle:2016dyj} and Belle~II~\cite{PhysRevD.110.072020}, and detector effects are emulated by event-by-event smearing of reconstructed momenta according to Belle~II tracking performance~\cite{Bertacchi_2021}.

We model $R(D)$ and $R(D^{*})$ in leptonic $\tautomuBelle$ decays and $R(D^{*})$ in one-prong hadronic $\tautohadBelle$ decays. All channels use the standard hadronic-tag discriminants $(m^2_{\rm miss},\,E_{\rm extra})$, where $m^2_{\rm miss}$ is the recoil mass squared computed in the center-of-mass frame under the back-to-back production hypothesis, where the energy of the accompanying $B$ meson in the event $E_{B_{\rm tag}}$ is equal to the beam energy in the same frame, corresponding to half of the total collision energy ~\cite{basf2-zenodo}, and $E_{\rm extra}$ is the calorimeter energy not assigned to either reconstructed $B$ meson. The templates are smeared to reproduce the resolution effects seen in public Belle~II results.

\begin{figure*}[!t]
  \centering
  \setlength{\tabcolsep}{4pt}
  \renewcommand{\arraystretch}{1.0}
  \vspace{-1mm}

  \begin{tabular}{c c c c c}
    & \multicolumn{2}{c}{\textbf{$B^0$}} & \multicolumn{2}{c}{\textbf{$B^+$}} \\
    & \textbf{$m^2_{\rm miss}$} & \textbf{$E_{\rm extra}$} & \textbf{$m^2_{\rm miss}$} & \textbf{$E_{\rm extra}$} \\
    \multicolumn{5}{c}{\rule{0pt}{2.2ex}} \\[-0.8ex]

    \rotatebox{90}{\hspace{1.4em}\textbf{$R(D^{*})$, $\tau\!\to\!\pi\nu$}} &
    \includegraphics[width=0.22\textwidth]{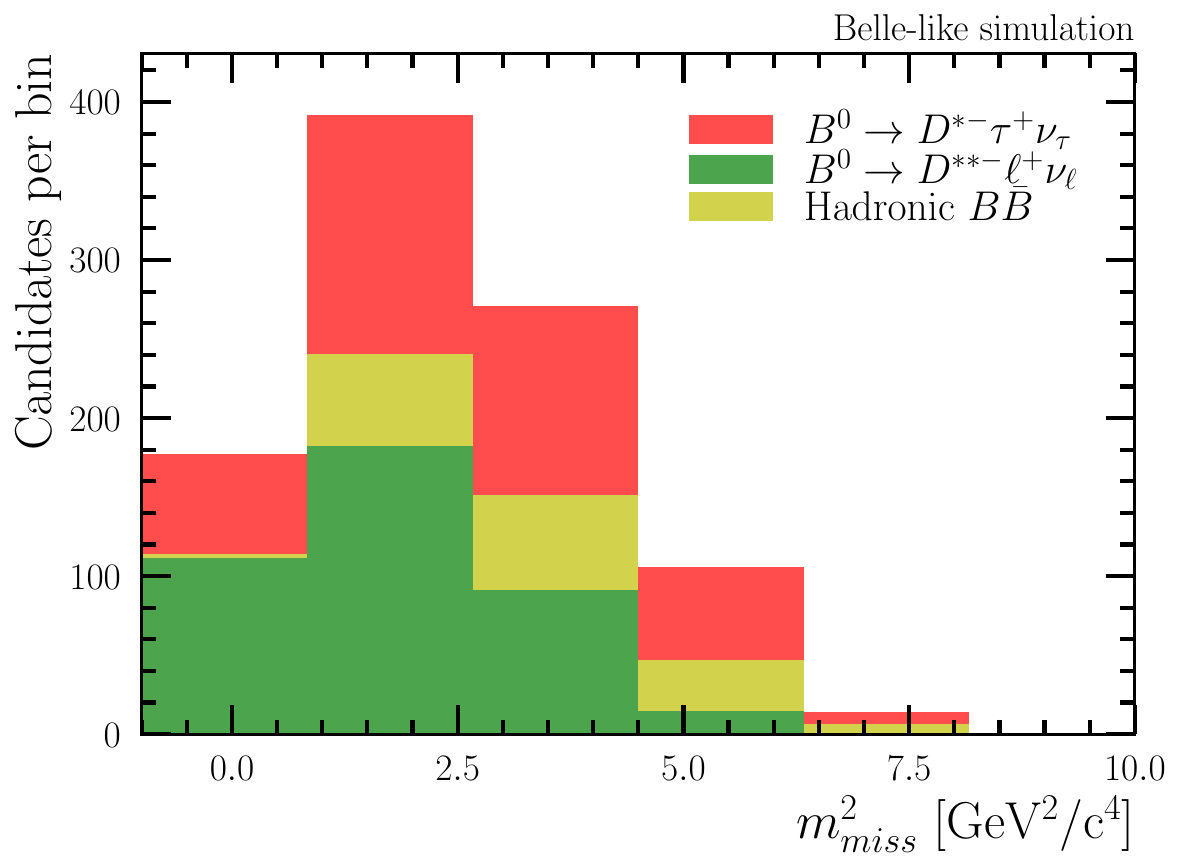} &
    \includegraphics[width=0.22\textwidth]{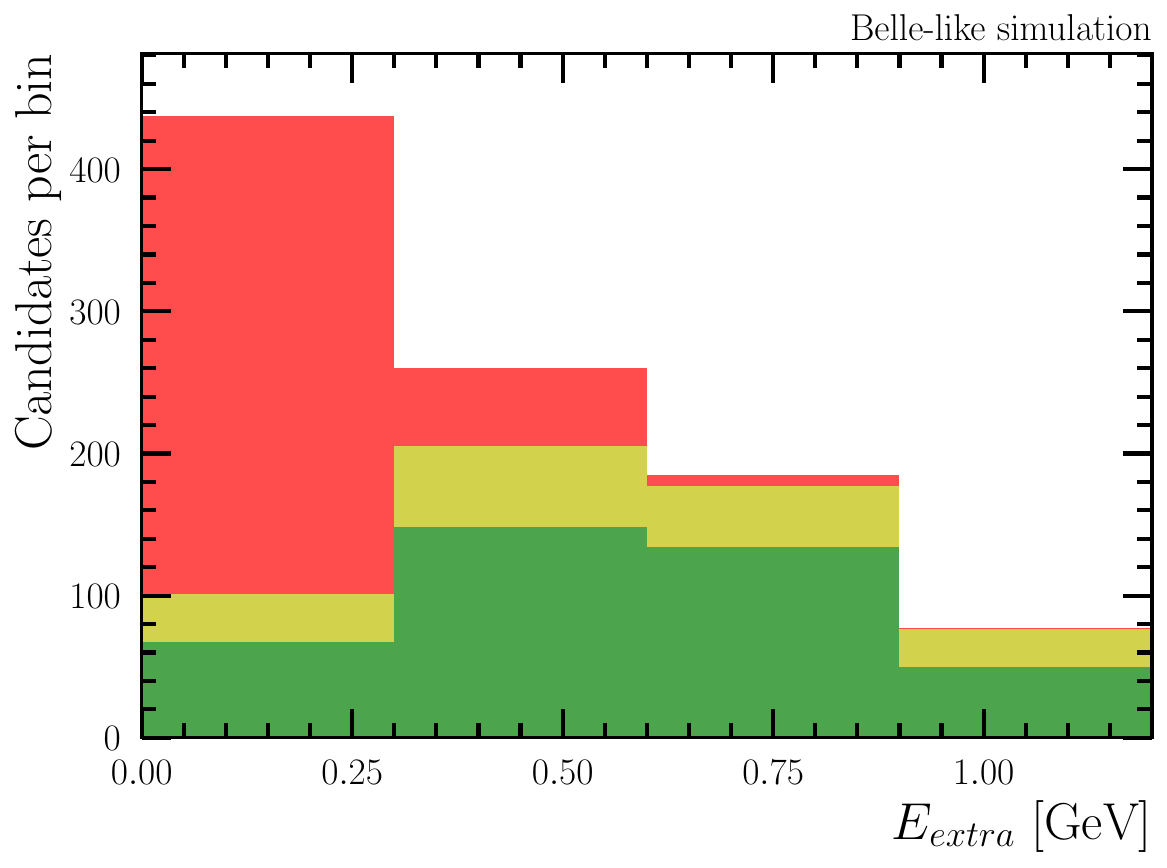} &
    \includegraphics[width=0.22\textwidth]{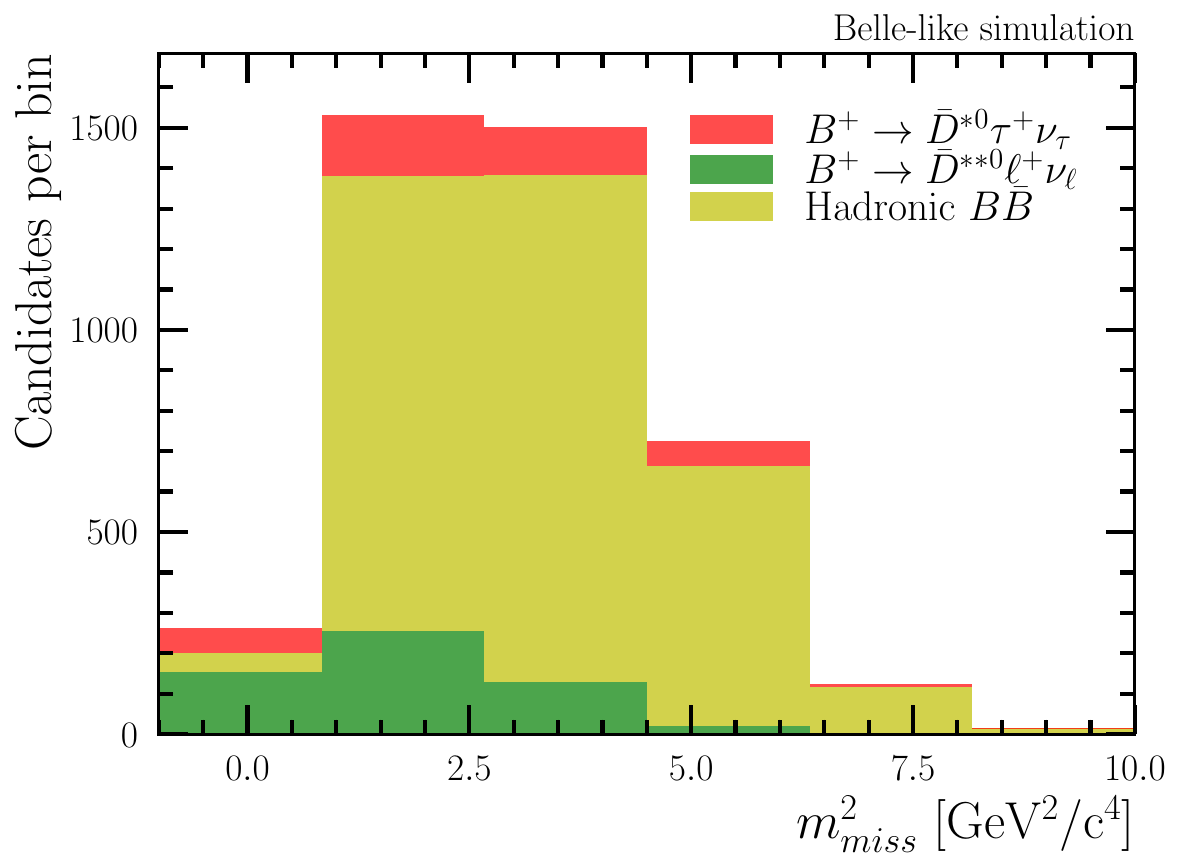} &
    \includegraphics[width=0.22\textwidth]{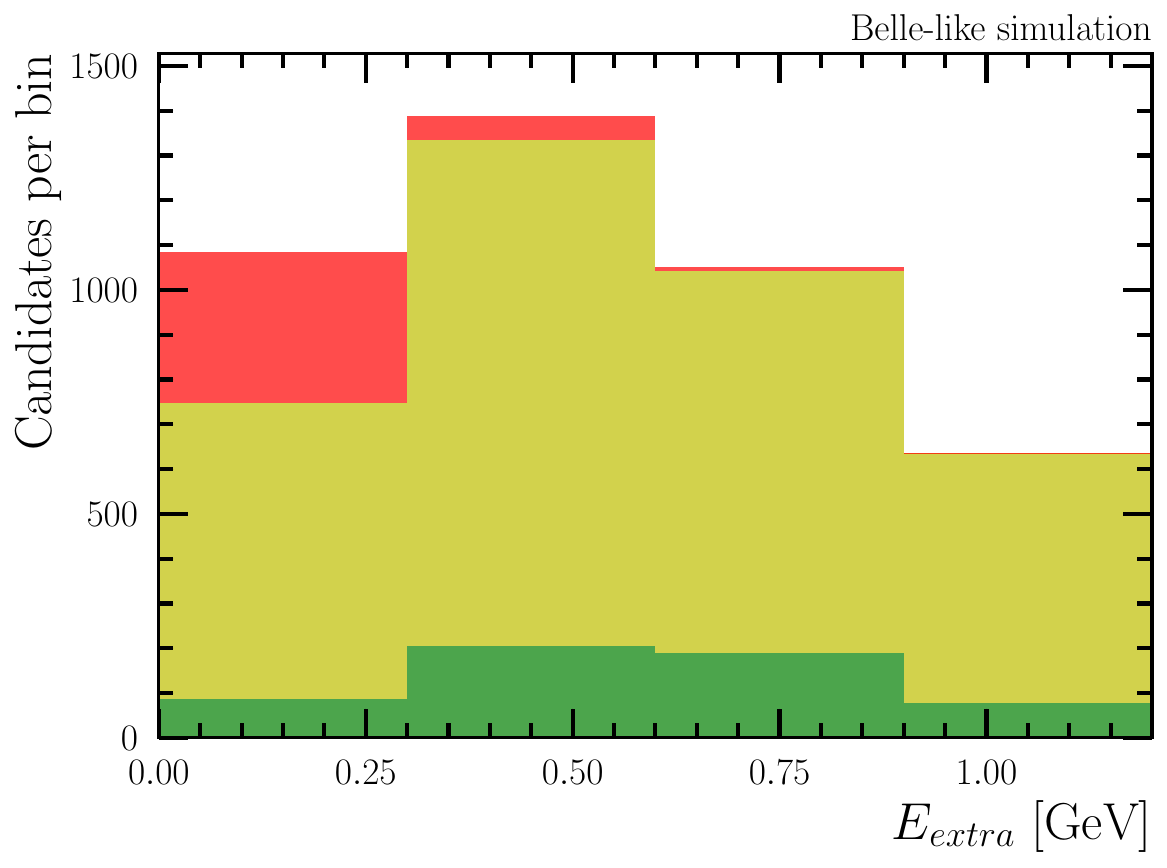} \\[0.7ex]

    \rotatebox{90}{\hspace{0.7em}\textbf{$R(D^{*})$, $\tau\!\to\!\ell\nu\nu$}} &
    \includegraphics[width=0.22\textwidth]{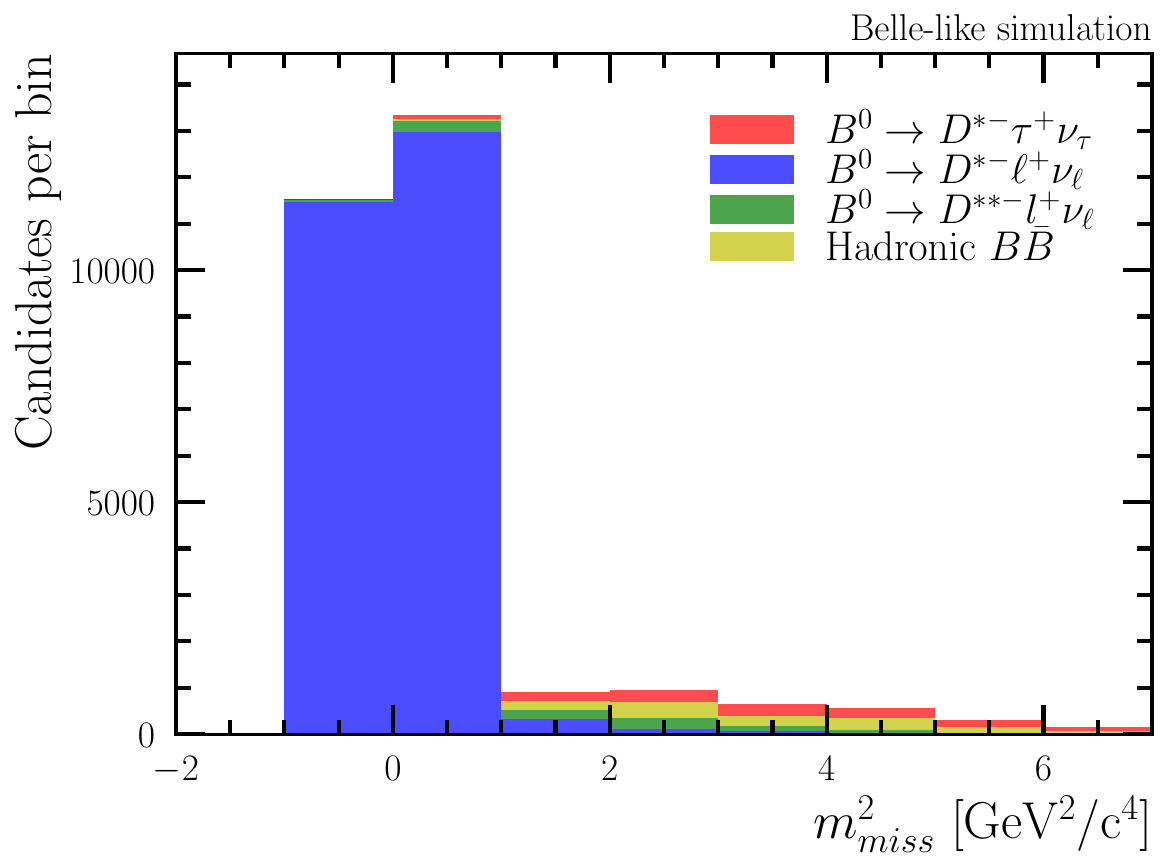} &
    \includegraphics[width=0.22\textwidth]{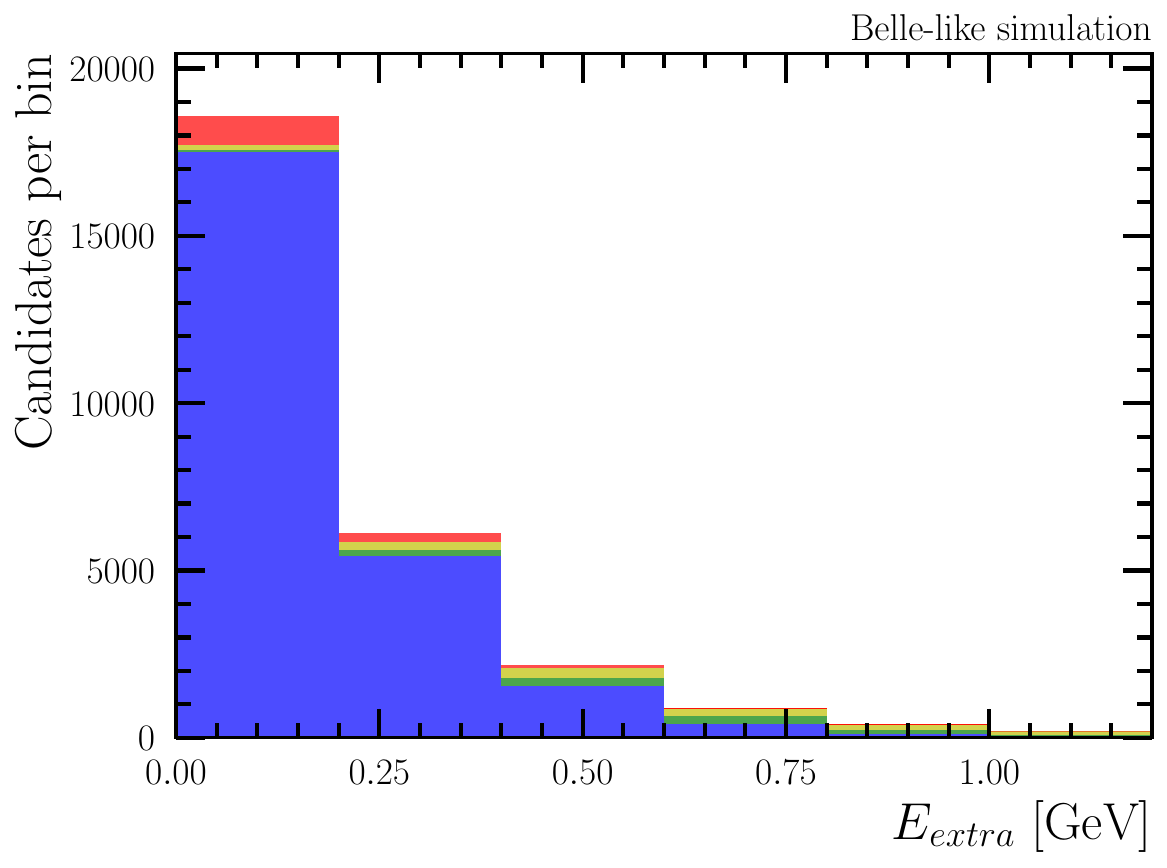} &
    \includegraphics[width=0.22\textwidth]{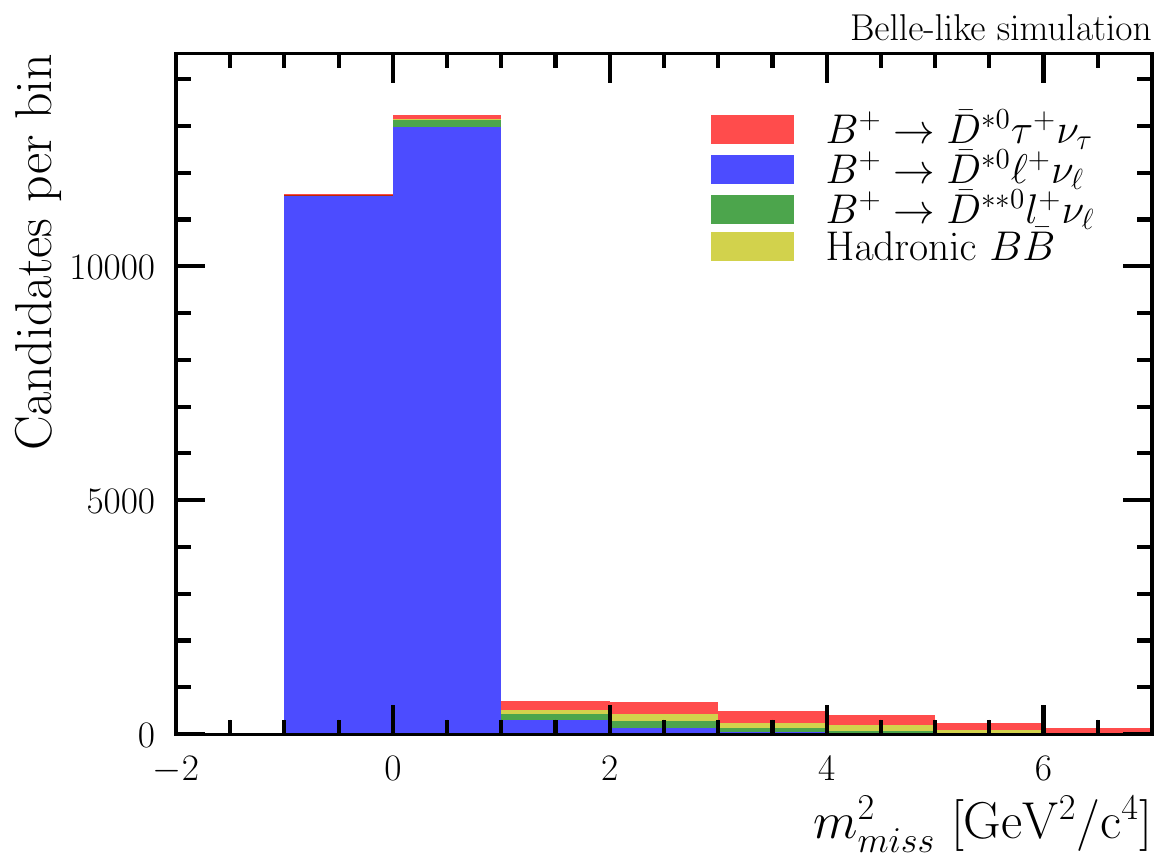} &
    \includegraphics[width=0.22\textwidth]{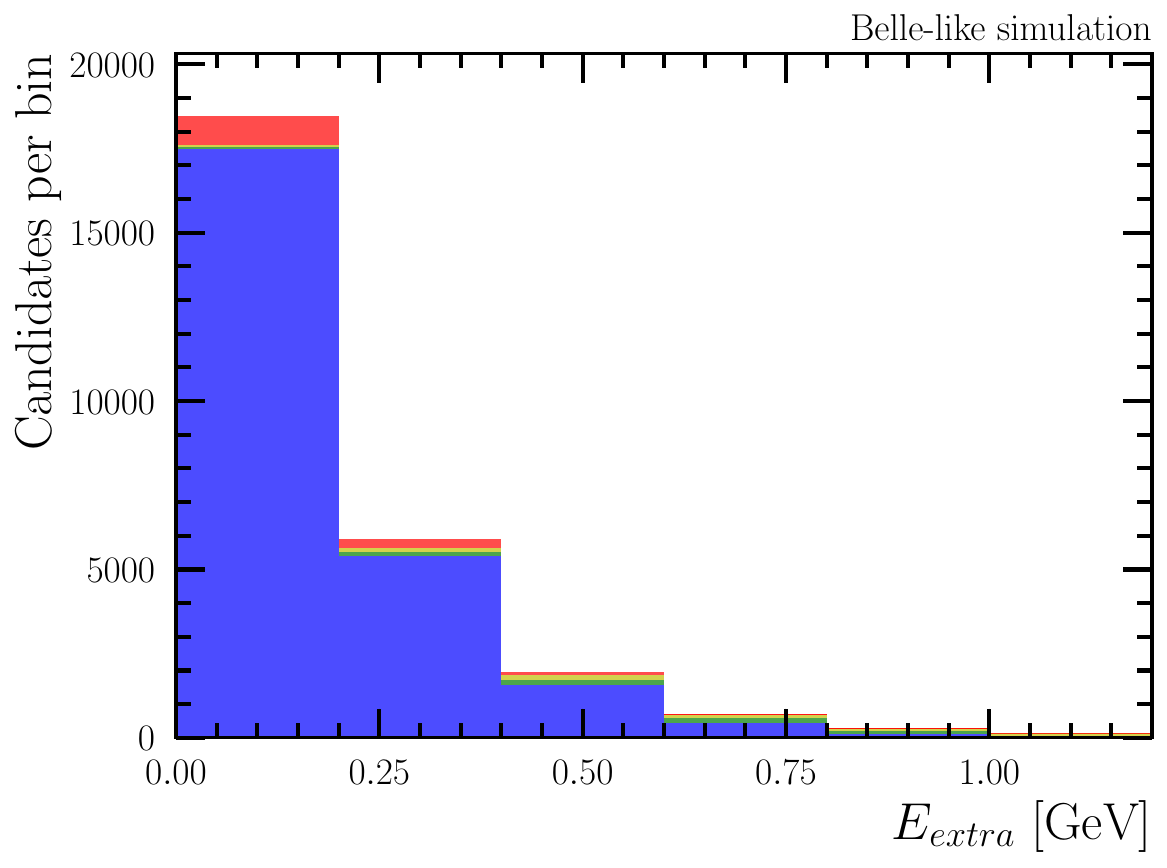} \\[0.7ex]

    \rotatebox{90}{\hspace{1.2em}\textbf{$R(D)$, $\tau\!\to\!\ell\nu\nu$}} &
    \includegraphics[width=0.22\textwidth]{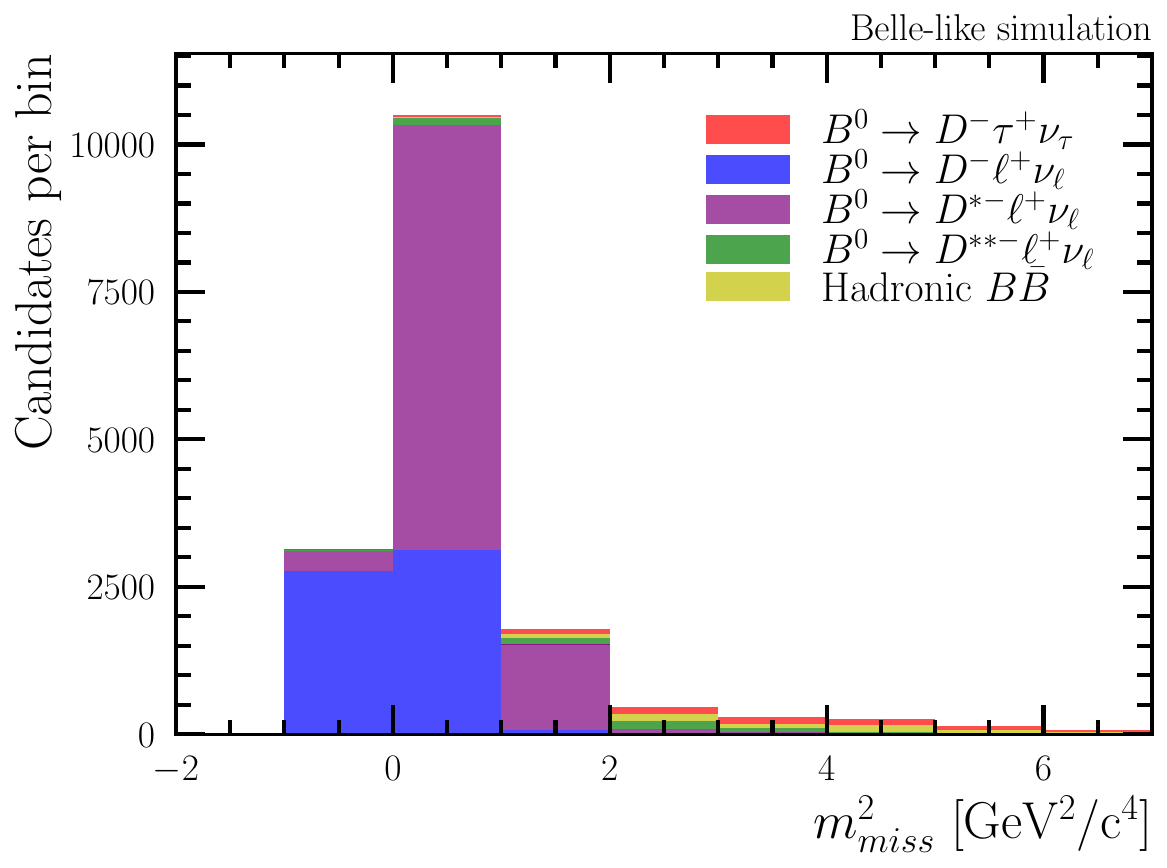} &
    \includegraphics[width=0.22\textwidth]{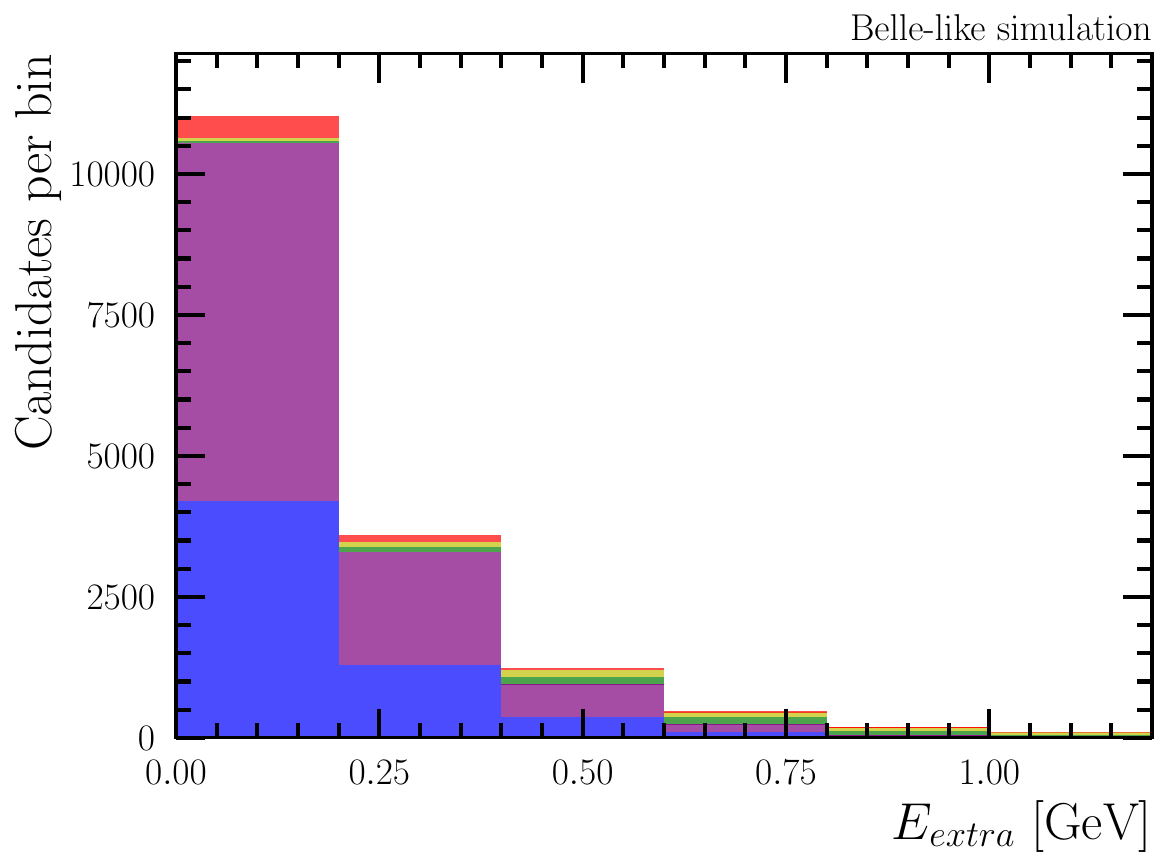} &
    \includegraphics[width=0.22\textwidth]{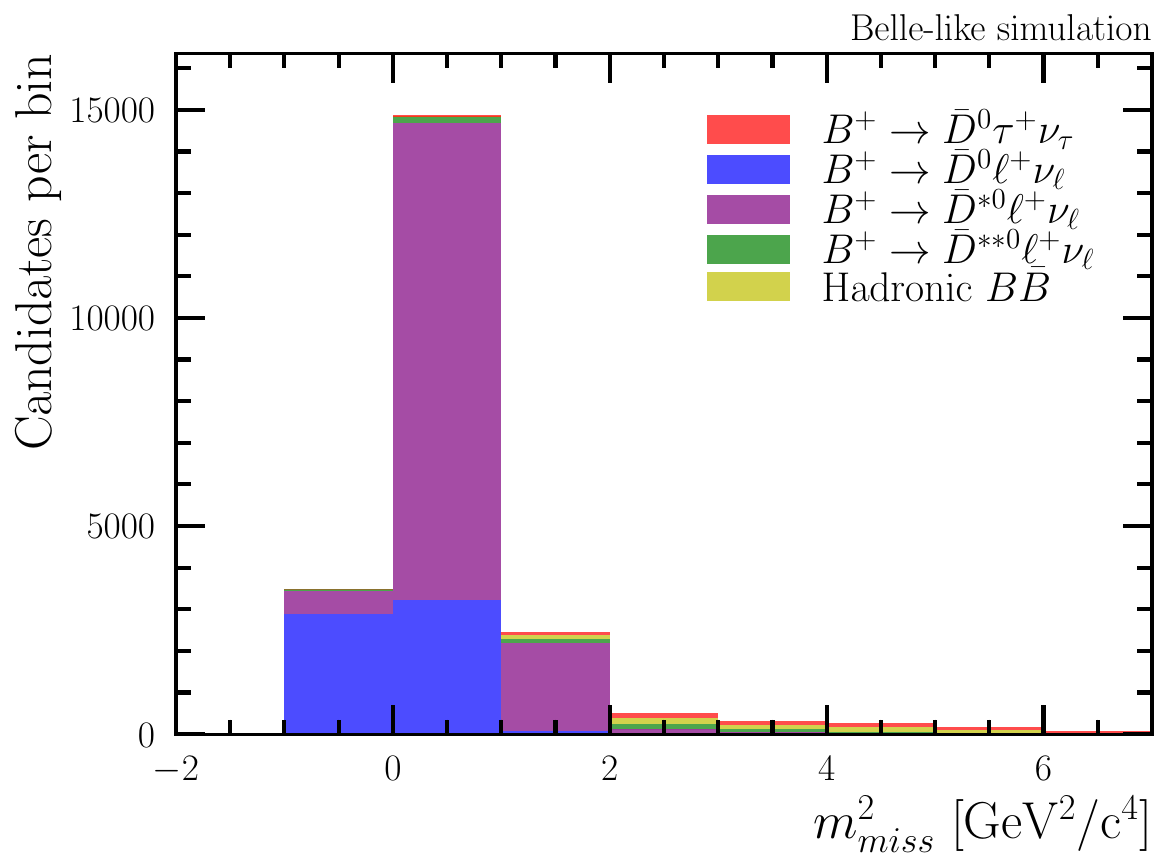} &
    \includegraphics[width=0.22\textwidth]{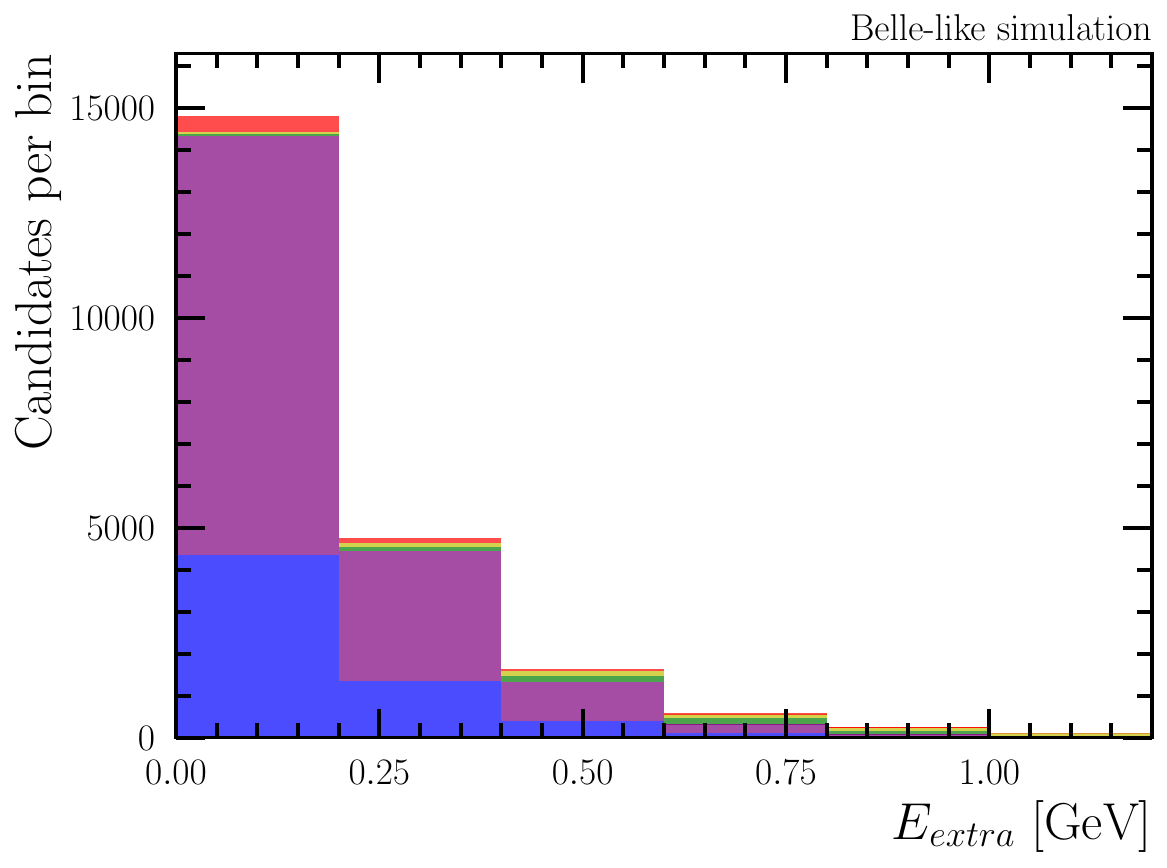} \\
  \end{tabular}

  \caption{Belle~II-like template projections arranged by category (rows) and $B$ charge (column blocks). Columns show the standard hadronic-tag observables $(m^2_{\rm miss}, E_{\rm extra})$ separately for $B^0$ and $B^+$. The $\tau\to\pi\nu$ row represents a one-prong hadronic category used in the $R(D^{*})$ configuration (see text). Signal is shown in red; other colours indicate background components.}
  \label{fig:BelleAllPanels}
\end{figure*}

A subset of the $\tau$-muonic simulation with $m_{\rm miss}<1~\mathrm{GeV}$ is used as a normalisation reference for the one-prong hadronic ($\tautohadBelle$) category. The resulting Belle~II-like template projections are shown in Fig.~\ref{fig:BelleAllPanels}, arranged so that the $B^0$ and $B^+$ contributions appear side-by-side for each category.

%% file: sec5.tex
\section{Combined sensitivity to Wilson coefficients}
\label{sec:sec5}

With the LHCb-like and Belle~II-like likelihood models defined in Section~\ref{sec:sec4}, we study the prospective constraints on the WET coefficient shifts
$\vec C=(C_{V_L},C_{V_R},C_{S_L},C_{S_R},C_T)$ from $B\to D^{(*)}\tau\nu$ measurements.
The goal is not a numerical ``average'' of inputs, but a controlled \emph{joint inference} in which (i) the short-distance parameters are common to all channels and experiments, and (ii) shared hadronic inputs--in particular the $B\to D^{(*)}$ form-factor parameters--enter as \emph{common nuisance parameters} within a single likelihood. This provides a well-defined global nuisance-parameter manifold and prevents inconsistencies that can arise when likelihoods are profiled independently and only combined afterwards.

While the framework supports simultaneous scans of the full coefficient vector, we focus on three representative directions: $C_{S_L}$ (scalar), $C_{V_R}$ (right-handed vector), and $C_T$ (tensor). These probe distinct helicity structures and compactly illustrate (i) channel complementarity across $D$ and $D^*$ final states and different $\tau$ decay modes, (ii) the impact of profiling shared hadronic degrees of freedom coherently across channels, and (iii) how inconsistent hadronic modelling can bias post-fit combinations. The remaining coefficients, $C_{V_L}$ and $C_{S_R}$, are not scanned in two dimensions here; for completeness, we quote conditional one-dimensional intervals for ${\rm Re}(C_{V_L})$ in Subsection~\ref{sec:sec5:future}.

\subsection{Asimov setup, benchmark precision, and scan procedure}
\label{sec:sec5:setup}

We generate \textit{Asimov} datasets following the standard prescription~\cite{Cowan:2011an} using the template models of Section~\ref{sec:sec4}.
Two injection hypotheses are considered:
\begin{itemize}
  \item \textbf{SM injection:} $\theta_0=(\vec C=\vec 0,\vec\alpha=\vec\alpha_0)$;
  \item \textbf{NP injection:} $\theta_0=(\vec C=\vec C_{\text{global-fit}},\vec\alpha=\vec\alpha_0)$, where $\vec C_{\text{global-fit}}$ is taken from a representative current global interpretation for each single Wilson Coefficient scenario we consider~\cite{Iguro:2024hyk}.
\end{itemize}
In both cases the reconstructed templates depend explicitly on $\theta=(\vec C,\vec\alpha)$ through the \textsc{ReDist}--\textsc{HAMMER} morphing described in Section~\ref{sec:sec3}, so variations of Wilson coefficients and form factors propagate coherently through acceptance, resolution, and template shapes.

\paragraph{Benchmark precision.}
To emulate prospective experimental reach, we set effective sample sizes such that the overall fractional constraint matches projected uncertainties on $R(D^{(*)})$ for the 2030 and 2040 benchmark scenarios of Refs.~\cite{Sevilla_2022,ATLAS:2025lrr}.
\begin{table}[t]
\centering
\small
\setlength{\tabcolsep}{5pt}
\renewcommand{\arraystretch}{1.12}
\caption{Benchmark relative uncertainties used to set the effective sample sizes in the sensitivity studies, based on the 2030 and 2040 projections of Refs.~\cite{Sevilla_2022,ATLAS:2025lrr}.}
\label{tab:stat}
\begin{tabular}{lcc}
\toprule
\textbf{Sample} & \textbf{2030} & \textbf{2040} \\
\midrule
LHCb-like $\BtoDst$, $\tautomuLHCb$ & $3.2\%$ & $3.0\%$ \\
LHCb-like $\BtoD$,   $\tautomuLHCb$ & $4.4\%$ & $3.3\%$ \\
LHCb-like $\BtoDst$, $\tautohadLHCb$   & $3.2\%$ & $3.0\%$ \\
Belle~II-like $\BtoDst$, $\tautomuBelle$ & $2.0\%$ & $1.0\%$ \\
Belle~II-like $\BtoD$,   $\tautomuBelle$ & $3.0\%$ & $1.4\%$ \\
Belle~II-like $\BtoDst$, $\tautohadBelle$      & $4.0\%$ & $2.0\%$ \\
\bottomrule
\end{tabular}
\end{table}
Operationally, we choose an effective signal yield $N_{\rm sig}$ such that
\begin{equation}
\frac{1}{\sqrt{N_{\rm sig}}}\ \sim\ \frac{\sigma_{R(D^{(*)})}}{R(D^{(*)})}\,,
\end{equation}
where $\sigma_{R(D^{(*)})}/R(D^{(*)})$ includes statistical and systematic components. In this way, $N_{\rm sig}$ should be interpreted as an \emph{effective constraining power} of the binned-template likelihood rather than a literal event count. The benchmark relative uncertainties used are summarized in Table~\ref{tab:stat}; the interpretation of this mapping, especially in a systematics-dominated regime, is discussed in Subsection~\ref{sec:sec5:implications}.

\paragraph{Scan definition and profiling.}
For each coefficient under study we scan the complex plane of its real and imaginary parts. At each grid point the scanned coefficient is fixed, while the nuisance parameters are profiled by minimizing the profile-likelihood ratio~\cite{Wilks:1938dza,Cowan:2011an}. The nuisance set includes the relevant $B\to D^{(*)}$ form-factor parameters $\vec\alpha$ and background normalizations; the $\BtoDststellnu$ feed-down form-factor parameters are constrained with Gaussian priors~\cite{Bernlochner:2017jxt} and shared across the LHCb-like channels (Section~\ref{sec:sec4:lhcb}). For SM-injected scans, Wilson coefficients not under study are fixed to zero. For each NP-injected scan only the relative Wilson coefficient is injected, while the other coefficients fixed to zero.
The Standard Model Wilson coefficient, shared between all the semileptonic modes, is fixed to the expected value as it is, in practice, very well constrained by the many modes that would be scaled by its variation.
Unless stated otherwise, $68\%$ confidence regions correspond to $\Delta(-2\ln\mathcal{L})=2.30$ for two scanned degrees of freedom, using the standard asymptotic mapping~\cite{Wilks:1938dza,PDG:2024}.

\paragraph{Two combination strategies.}
We compare:
\begin{itemize}
  \item \textbf{Combined fit:} a single likelihood in which the Wilson coefficients and the shared form-factor parameters are common across all contributing channels;
  \item \textbf{Post-fit sum:} a point-by-point sum of single-channel \emph{profiled} likelihoods, where nuisance parameters are profiled independently in each channel prior to summation.
\end{itemize}
The two approaches coincide only if profiling is performed over an identical nuisance-parameter manifold in each channel. When channels are allowed to ``profile away'' discrepancies using different effective nuisance directions, the post-fit sum need not correspond to any globally consistent likelihood.

\subsection{Channel complementarity and physics interpretation}
\label{sec:sec5:results}

\begin{figure*}[!htp]
\centering
\begin{subfigure}[t]{0.49\textwidth}
  \centering
  \includegraphics[width=\linewidth]{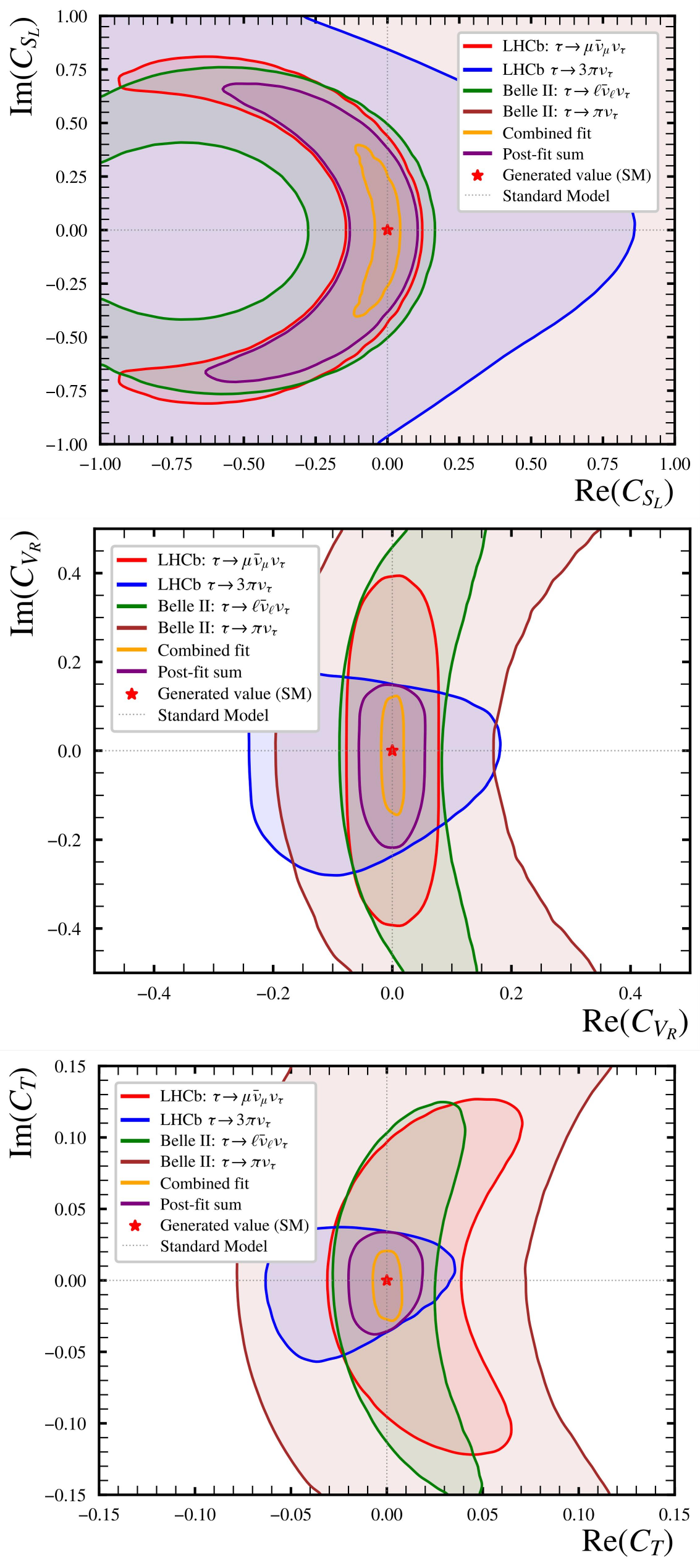}
  \caption{SM injection ($\vec C=\vec 0$).}
\end{subfigure}\hfill
\begin{subfigure}[t]{0.49\textwidth}
  \centering
  \includegraphics[width=\linewidth]{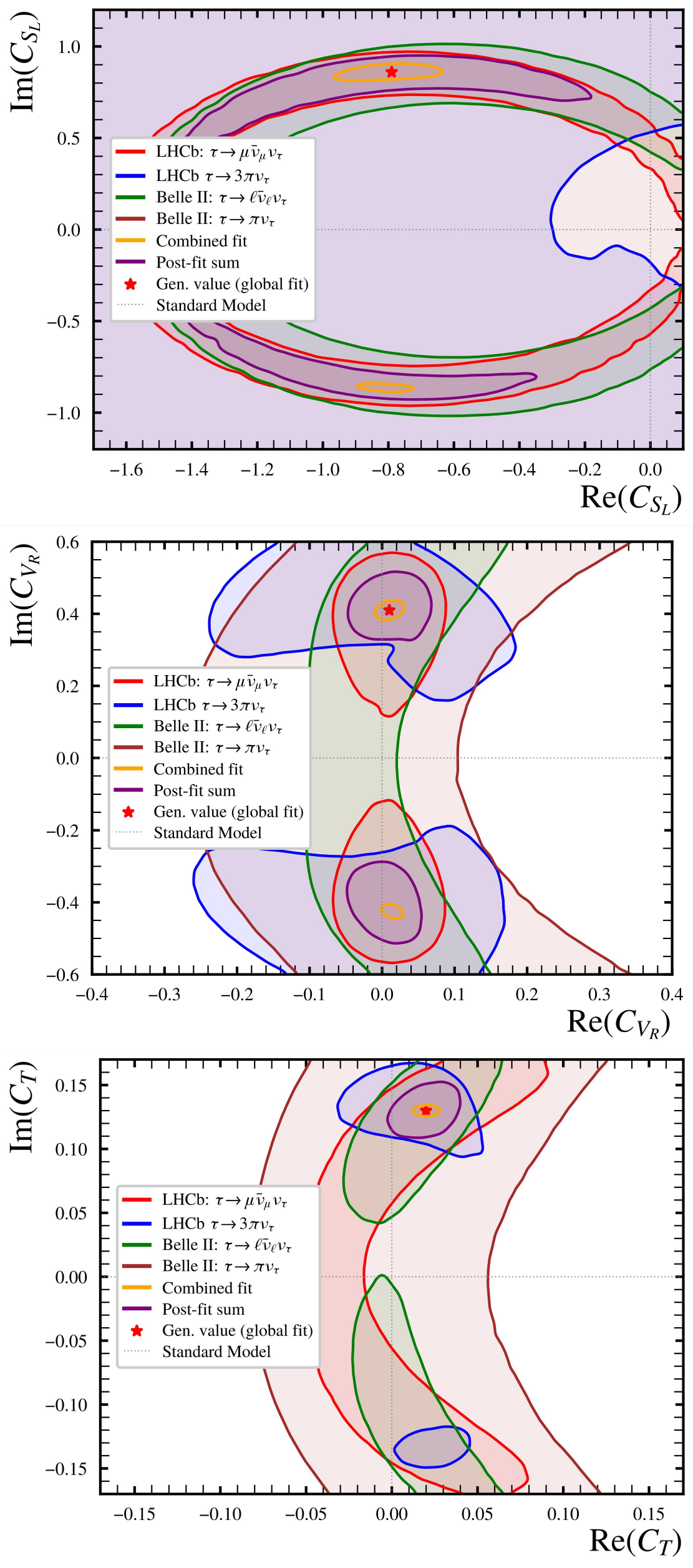}
  \caption{NP injection ($\vec C=\vec C_{\text{global-fit}}$)~\cite{Iguro:2024hyk}.}
\end{subfigure}
\caption{$68\%$ confidence regions in the complex planes of (top) $C_{S_L}$, (middle) $C_{V_R}$, and (bottom) $C_T$ for the 2030 benchmark.
The star indicates the injected point; the SM point at $(0,0)$ is shown for reference in the NP-injected case.}
\label{fig:scan_2030}
\end{figure*}

Figure~\ref{fig:scan_2030} shows $68\%$ confidence regions in the complex planes of $C_{S_L}$, $C_{V_R}$ and $C_T$ for the 2030 benchmark, for both SM and NP injections.
The individual categories yield contours with different orientations because they probe different combinations of helicity amplitudes through distinct reconstructed observables, and because their residual degeneracies with profiled form-factor parameters are not identical. The combined sensitivity is therefore driven primarily by the intersection of complementary degeneracy directions across channels, rather than by any single rate-dominated measurement.

\paragraph{Interference pattern and phases.}
For CP-averaged observables and in the absence of sizeable strong phases, the leading dependence is typically linear in ${\rm Re}(C)$ through SM--NP interference, while ${\rm Im}(C)$ contributes mainly through quadratic terms such as $|C|^2$. This generically implies stronger reach on real parts and an approximate symmetry under ${\rm Im}(C)\to-{\rm Im}(C)$, which can lead to disconnected allowed regions in NP-injected scans. Differential (shape) information and correlations with profiled hadronic parameters can partially break this symmetry, but it is not fully removed without genuinely CP-odd observables.

\paragraph{$C_{S_L}$ (scalar): primary sensitivity from $\BtoD$.}
Scalar currents modify $\BtoDtaunu$ most directly, so categories with substantial $B\to D$ sensitivity provide the dominant handle. In the SM-injected scan, rate information alone would admit an approximately annular locus in $(\mathrm{Re},\mathrm{Im})$ space; including shape information and profiling over $\vec\alpha$ contracts this into the observed crescent-like region. In the NP-injected case, the preferred region moves to the vicinity of the injected point, with any secondary lobe reflecting the approximate phase-reflection symmetry of CP-averaged information.

\paragraph{$C_{V_R}$ (right-handed vector): enhanced discrimination in $B\to D^*$.}
Sensitivity to $C_{V_R}$ is suppressed for pseudoscalar final states but enhanced in $B\to D^*$, where the vector meson provides additional helicity structure. The resulting contours rotate between $D$- and $D^*$-dominated categories, and channels with greater helicity and $\tau$-polarisation sensitivity supply directions that localise the solution when combined.

\paragraph{$C_T$ (tensor): strong shape leverage in $\BtoDst$.}
Tensor interactions produce pronounced distortions of the differential decay structure in $\BtoDsttaunu$, leading to characteristic template deformations in reconstructed observables. This typically generates more orthogonal degeneracy directions between categories and makes the tensor direction especially responsive to multi-channel combinations.

\paragraph{Role of shared hadronic inputs.}
Across all three operators, constraints tighten when the dominant form-factor degrees of freedom are treated consistently across categories. Channels with strong normalisation and form-factor handles calibrate nuisance directions that can otherwise mimic NP-like deformations elsewhere through their correlation with Wilson Coefficients parameters. Sharing the form-factor parameters improves the localisation of the short-distance parameters in a combined fit as it includes additional constraints to the parameters across all of the different modes that, in single mode scenarios, constraint respectively different regions of the phase space of the form-factors.

\subsection{Bias from inconsistent hadronic models in post-fit combinations}
\label{sec:sec5:bias}

Beyond a loss of sensitivity, a post-fit sum can be biased if different channels are interpreted with different hadronic parameterisations. Even if each channel is internally consistent, summing \emph{profiled} likelihoods defined on different nuisance manifolds (e.g.\ different truncations or constraint schemes in dispersive expansions) does not, in general, correspond to profiling a single globally consistent physical model. The problem is most acute when the alternative parameterisations admit different effective shape-deformation directions, so that ``profiling away'' a discrepancy corresponds to different physical template variations in different channels~\cite{Boyd:1995sq,Boyd:1997kz,Bernlochner:2022hqe}.

\begin{figure}[t]
\centering
\includegraphics[width=\linewidth]{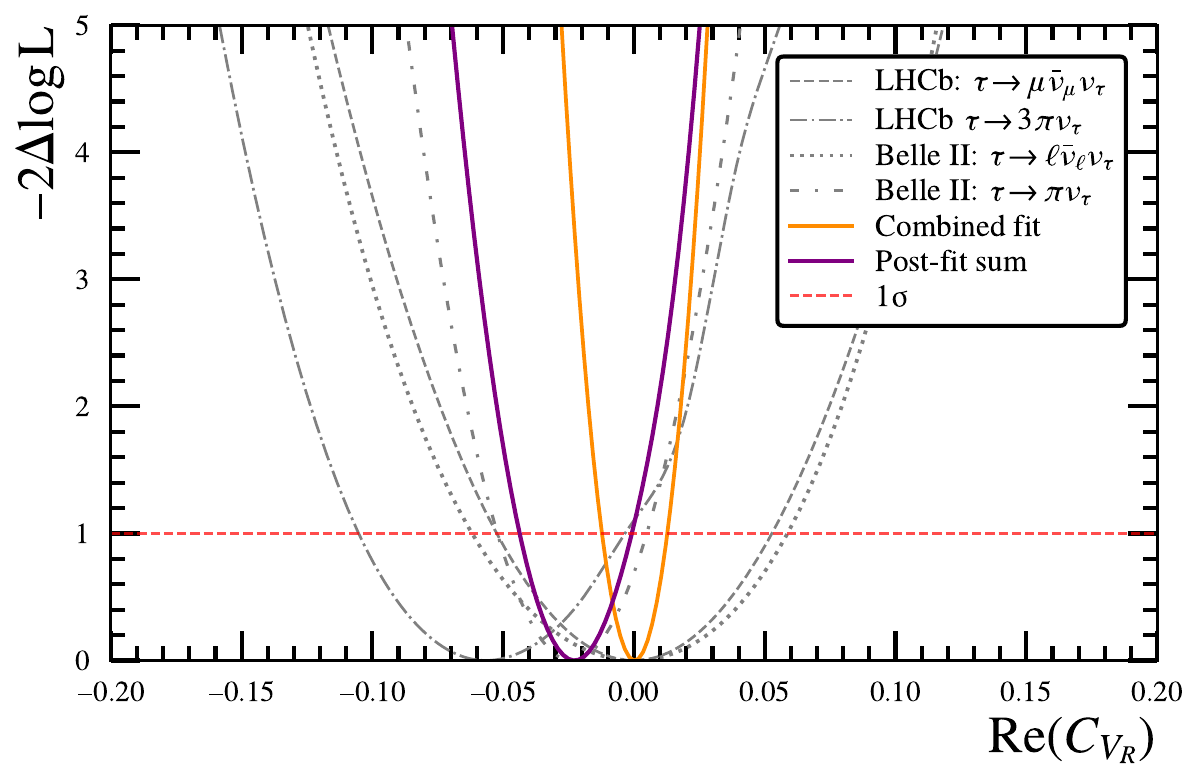}
\caption{One-dimensional likelihood scans in ${\rm Re}(C_{V_R})$ illustrating bias from inconsistent hadronic modelling in a post-fit combination.
The Asimov dataset is generated using BLPRXP~\cite{Bernlochner:2022hqe}, while in the interpretation step the hadronic-$\tau$ categories are scanned using BGL~\cite{Boyd:1995sq,Boyd:1997kz}.
The individual channel scans develop misaligned minima and the post-fit sum inherits a displaced minimum.
A unified likelihood-level treatment in a common parameterisation enforces consistent profiling of the shared hadronic degrees of freedom and removes this artefact.}
\label{fig:show_bias}
\end{figure}

Figure~\ref{fig:show_bias} illustrates this with a toy example. An SM-injected Asimov dataset is generated using the nominal BLPRXP form-factor model~\cite{Bernlochner:2022hqe}. In the \emph{interpretation} step, the hadronic-$\tau$ categories are instead described with a BGL parameterisation~\cite{Boyd:1995sq,Boyd:1997kz}, while the remaining channels retain BLPRXP. Residual mismodelling that cannot be absorbed by profiling then pulls the best-fit short-distance hypothesis; in this example ${\rm Re}(C_{V_R})$ provides an efficient lever arm through its impact on helicity composition and $q^2$ dependence. The post-fit sum inherits the resulting misalignment as an apparent displacement of the combined minimum. A unified likelihood-level treatment avoids this artefact by enforcing a common hadronic basis and profiling a shared set of hadronic degrees of freedom for each Wilson-coefficient hypothesis.

\subsection{Evolution with benchmark precision}
\label{sec:sec5:future}

\begin{figure*}[!htp]
\centering
\begin{subfigure}[t]{0.49\textwidth}
  \centering
  \includegraphics[width=\linewidth]{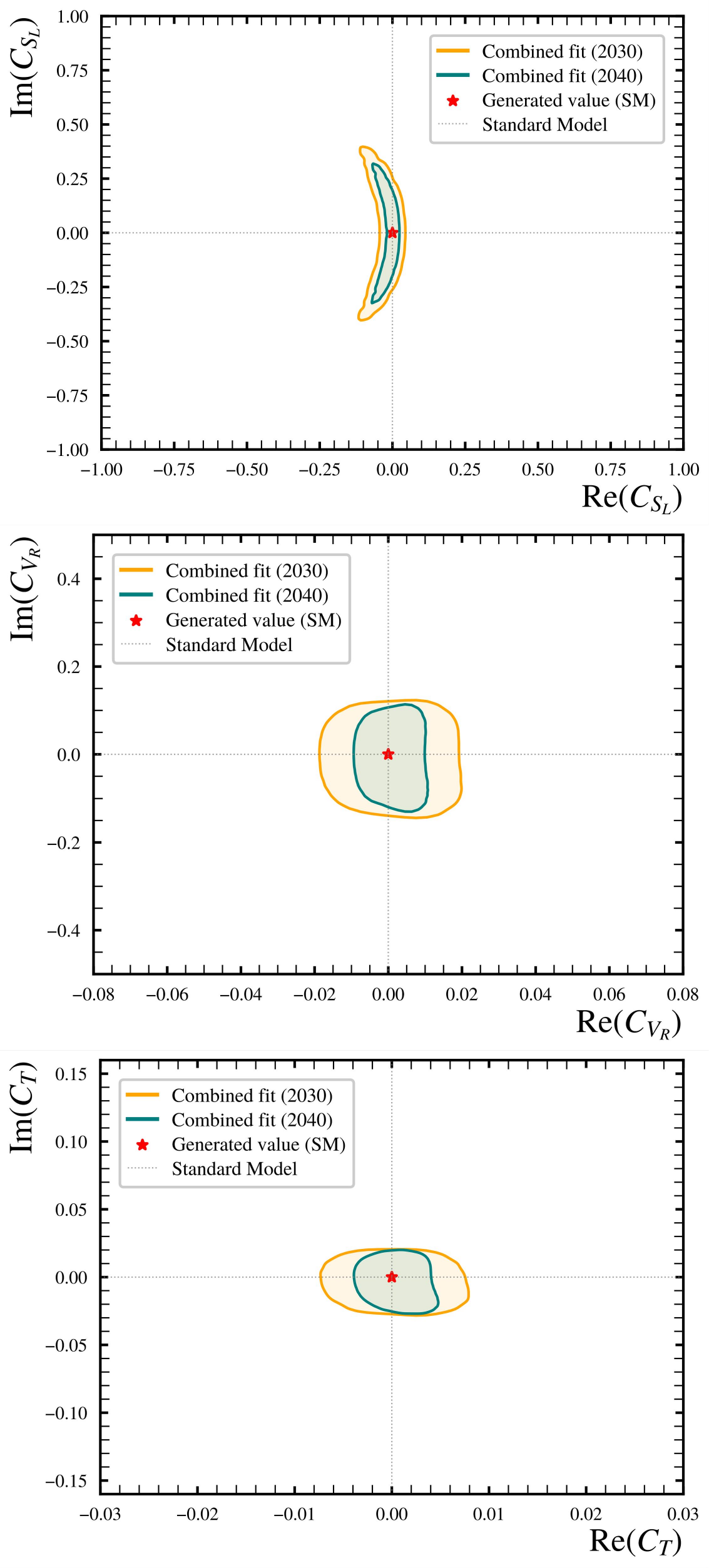}
  \caption{SM-injected Asimov dataset ($\vec C=\vec 0$): combined-fit projections for 2030 and 2040.}
\end{subfigure}\hfill
\begin{subfigure}[t]{0.49\textwidth}
  \centering
  \includegraphics[width=\linewidth]{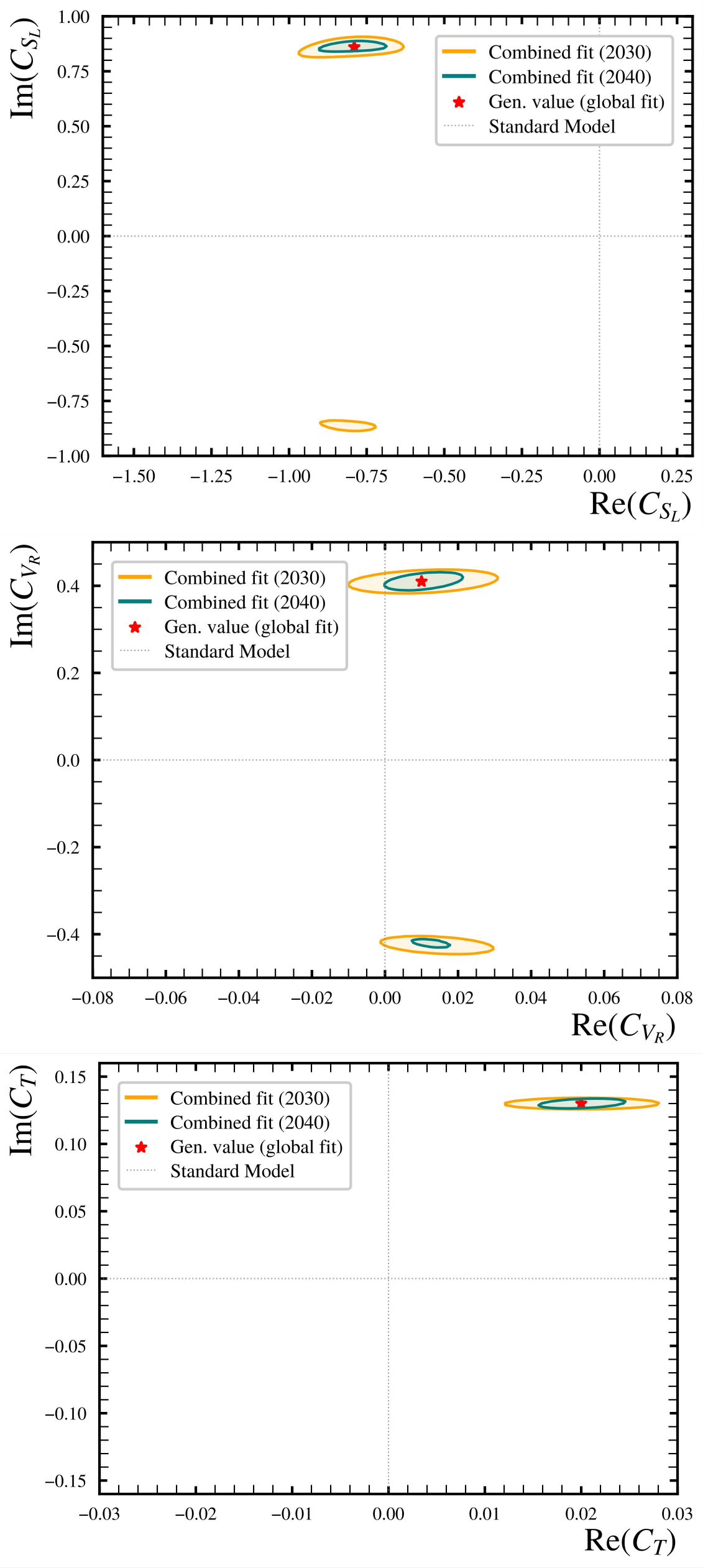}
  \caption{NP-injected Asimov dataset ($\vec C=\vec C_{\text{global-fit}}$)~\cite{Iguro:2024hyk}: combined-fit projections for 2030 and 2040.}
\end{subfigure}
\caption{Projected combined-fit $68\%$ confidence regions for the 2030 (orange) and 2040 (teal) benchmark scenarios in the complex planes of
(left) $C_{S_L}$, (middle) $C_{V_R}$, and (right) $C_T$.
The star indicates the injected point; the SM point at $(0,0)$ is shown for orientation in the NP-injected case.}
\label{fig:benchmarks}
\end{figure*}

We next study how the combined-fit sensitivity evolves under the projected 2030 and 2040 benchmark scenarios of Refs.~\cite{Sevilla_2022,ATLAS:2025lrr}, keeping the analysis model, nuisance parameterisation, and profiling strategy fixed.
Figure~\ref{fig:benchmarks} compares the combined-fit $68\%$ regions in the complex planes of $C_{S_L}$, $C_{V_R}$, and $C_T$.
Across coefficients, the 2040 contours contract relative to 2030 while preserving the characteristic geometry set by interference and nuisance profiling, indicating that in these simplified studies the dominant limiting directions are largely controlled by physics and shared-hadronic correlations rather than by pure counting statistics.

\paragraph{Conditional one-dimensional intervals.}
To complement the two-dimensional contours, Table~\ref{tab:wilson_1d_conditional} quotes \emph{conditional} one-dimensional $68\%$ CL intervals obtained by floating a single coefficient component (real or imaginary) while profiling nuisance parameters. For SM-injected datasets, all other Wilson coefficients are fixed to zero; for NP-injected datasets, all the coefficients of interest of each scenario are fixed to their injected values, while the others are fixed to zero, so the intervals quantify the conditional local precision around the benchmark point. These should not be interpreted as marginalised constraints from a simultaneous multi-parameter EFT fit. In particular, with CP-averaged information the likelihood can exhibit approximate phase-reflection symmetries in imaginary components, which can lead to unions of disconnected intervals; robust removal of such ambiguities ultimately requires CP-odd observables (e.g.\ CP asymmetries or signed triple-product correlations), which are beyond the scope of this paper.

\begin{table*}[!htp]
\centering
\small
\setlength{\tabcolsep}{4.6pt}
\renewcommand{\arraystretch}{1.20}
\caption{Conditional one-dimensional $68\%$ CL intervals from Asimov datasets for two injection scenarios, shown in a single table to facilitate direct comparison.
Only the listed component is floated (other Wilson coefficients fixed as described in the text; nuisances profiled).
Results are quoted as $\pm x$ with $x=\max(|a|,|b|)$ for an interval $[a,b]$, except for sizeable asymmetry ($>30\%$), where $[a,b]$ is shown.
Only ${\rm Re}(C_{V_L})$ is reported, since ${\rm Im}(C_{V_L})$ enters CP-averaged observables only quadratically and is weakly constrained in this setup.
In the NP-injected panel, unions indicate disconnected solutions from an approximate phase-reflection ambiguity in CP-averaged information; this is most prominent in the Belle~II-like imaginary parts and is reduced by the inclusion of LHCb-like $\tau\to3\pi\nu$ information and by the combined fit.
NP-injected intervals are quoted relative to the injected global-fit point from Ref.~\cite{Iguro:2024hyk},
$(C_{V_L},C_{S_L},C_{V_R},C_T)_{\rm inj}=(0.079,\,-0.79+i\,0.86,\,0.01+i\,0.41,\,0.02+i\,0.13)$.}
\label{tab:wilson_1d_conditional}
\resizebox{\textwidth}{!}{%
\begin{tabular}{lcccccc}
\toprule
 & \multicolumn{3}{c}{\textbf{2030}} & \multicolumn{3}{c}{\textbf{2040}} \\
\cmidrule(lr){2-4}\cmidrule(lr){5-7}
\textbf{Parameter} & \textbf{LHCb-like} & \textbf{Belle~II-like} & \textbf{Combined} &
\textbf{LHCb-like} & \textbf{Belle~II-like} & \textbf{Combined} \\
\midrule
\multicolumn{7}{l}{\textbf{SM-injected Asimov datasets ($C_{\rm inj}=0$)}} \\
\midrule
${\rm Re}(C_{V_L})$ & $\pm 0.031$ & $\pm 0.055$ & $\pm 0.012$ &
$\pm 0.028$ & $\pm 0.023$ & $\pm 0.007$ \\
\addlinespace[2pt]
${\rm Re}(C_{S_L})$ & $\pm 0.11$ & $[-0.85,\,0.09]$ & $[-0.07,\,0.03]$ &
$\pm 0.10$ & $[-0.27,\,0.04]$ & $[-0.05,\,0.02]$ \\
${\rm Im}(C_{S_L})$ & $\pm 0.36$ & $\pm 0.67$ & $\pm 0.31$ &
$\pm 0.35$ & $\pm 0.58$ & $\pm 0.27$ \\
\addlinespace[2pt]
${\rm Re}(C_{V_R})$ & $\pm 0.030$ & $[-0.043,\,0.070]$ & $\pm 0.013$ &
$\pm 0.027$ & $[-0.016,\,0.030]$ & $\pm 0.007$ \\
${\rm Im}(C_{V_R})$ & $\pm 0.12$ & $\pm 0.43$ & $\pm 0.11$ &
$\pm 0.11$ & $\pm 0.28$ & $\pm 0.10$ \\
\addlinespace[2pt]
${\rm Re}(C_T)$     & $\pm 0.011$ & $\pm 0.022$ & $\pm 0.005$ &
$\pm 0.0095$ & $\pm 0.007$ & $[-0.002,\,0.003]$ \\
${\rm Im}(C_T)$     & $\pm 0.020$ & $\pm 0.10$ & $[-0.02,\,0.01]$ &
$\pm 0.020$ & $\pm 0.06$ & $[-0.02,\,0.01]$ \\
\midrule\midrule
\multicolumn{7}{l}{\textbf{NP-injected Asimov datasets (intervals relative to corresponent entries of $C_{\rm inj}$)}} \\
\midrule
${\rm Re}(C_{V_L})$ & $\pm 0.030$ & $\pm 0.058$ & $\pm 0.012$ &
$\pm 0.027$ & $\pm 0.024$ & $\pm 0.007$ \\
\addlinespace[2pt]
${\rm Re}(C_{S_L})$ & $\pm 0.16$ & $\pm 0.76$ & $\pm 0.11$ &
$\pm 0.15$ & $\pm 0.45$ & $\pm 0.08$ \\
${\rm Im}(C_{S_L})$ & $\pm 0.09$ &
$[-1.81,\,-1.51]\cup[-0.20,\,0.09]$ & $\pm 0.03$ &
$\pm 0.07$ & $[-0.08,\,0.04]$ & $\pm 0.02$ \\
\addlinespace[2pt]
${\rm Re}(C_{V_R})$ & $\pm 0.032$ & $\pm 0.071$ & $\pm 0.013$ &
$\pm 0.028$ & $\pm 0.030$ & $\pm 0.006$ \\
${\rm Im}(C_{V_R})$ & $[-0.02,\,0.01]$ &
$[-1.01,\,-0.49]\cup[-0.33,\,0.19]$ & $\pm 0.01$ &
$[-0.02,\,0.01]$ &
$[-0.90,\,-0.73]\cup[-0.10,\,0.09]$ & $\pm 0.01$ \\
\addlinespace[2pt]
${\rm Re}(C_T)$     & $\pm 0.013$ & $[-0.032,\,0.020]$ & $\pm 0.005$ &
$\pm 0.012$ & $[0.006,\,0.012]$ & $\pm 0.003$ \\
${\rm Im}(C_T)$     & $\pm0.003$ &
$[-0.282,\,-0.195]\cup[-0.039,\,0.032]$ & $\pm0.003$ &
$\pm 0.003$ & $\pm 0.016$ & $\pm0.003$ \\
\bottomrule
\end{tabular}
}
\end{table*}

\subsection{Benchmark interpretation and implications for joint LHCb--Belle~II fits}
\label{sec:sec5:implications}

The effective-yield normalisation used throughout this section should be read as an \emph{effective constraining power}, not as a literal event count. This distinction matters because many $B\to D^{(*)}\tau\nu$ measurements are already, or will soon become, systematics limited. By matching $\sigma_{R(D^{(*)})}/R(D^{(*)})$ we implicitly fold anticipated improvements in the dominant experimental limitations that govern the template likelihood (e.g.\ simulated-sample precision, feed-down and double-charm background control, and improved external constraints on form factors entering signal and background descriptions). The present study does not attempt to decompose systematics or model their bin-to-bin and channel-to-channel correlations; such a treatment requires the full experimental machinery, including control regions and an explicit nuisance-parameter construction. Accordingly, the contour sizes and one-dimensional intervals reported above should be interpreted as prospective EFT sensitivities \emph{conditional on achieving the projected total precision}.

Within that regime, the main added value of a joint LHCb--Belle~II interpretation is not simply more events, but the ability to represent \emph{shared physics inputs} coherently at the likelihood level. Detector- and environment-specific uncertainties (trigger and reconstruction at LHCb; tag-side efficiency and missing-energy modelling at Belle~II) are largely uncorrelated and combine straightforwardly once likelihoods are available. In contrast, theory-driven and physics-modelling uncertainties--most notably the $B\to D^{(*)}$ form-factor degrees of freedom (and common inputs entering feed-down and background descriptions)--are logically shared across measurements and should be represented by common nuisance parameters in a joint fit. This motivates publishing portable, morphable likelihoods in which the dominant shared hadronic parameters are explicit, enabling consistent EFT constraints without relying on post-fit sums of independently profiled likelihoods and without incurring the modelling-induced biases illustrated in Subsection~\ref{sec:sec5:bias}.

%% file: sec6.tex
\section{Summary and conclusions}
\label{sec:sec6}

Motivated by the persistent tension in $R(D^{(*)})$ and by the need for EFT interpretations that remain reliable away from the SM point, we have performed a first sensitivity study of a \emph{combined}, likelihood-level extraction of WET Wilson coefficients in $B\to D^{(*)}\tau\nu$ using simplified LHCb- and Belle~II-like template configurations. 
The study targets two challenges: the consistent propagation of shared hadronic inputs across measurements, and the robustness of template-based inferences when non-SM hypotheses coherently deform kinematic distributions, acceptance, and reconstructed shapes.

The results in Section~\ref{sec:sec5} demonstrate that a joint likelihood fit is not merely advantageous, but \emph{necessary} for a well-defined EFT interpretation. 
The improvement comes from profiling a \emph{common} set of $B\to D^{(*)}$ form-factor parameters across channels and experiments, ensuring that categories with pronounced hadronic sensitivity (and, where applicable, strong normalisation handles) constrain nuisance directions that would otherwise masquerade as short-distance effects. 
At the same time, the combined fit benefits of channel complementarity: constraints emerge from the intersection of degeneracy directions across $D$ and $D^*$ final states and across distinct $\tau$ decay modes, rather than from any single rate-dominated input. 
By contrast, post-fit summations of independently profiled likelihoods can forfeit this information flow and, more seriously, can become \emph{internally inconsistent} when channels are interpreted with different hadronic parameterisations or constraint schemes.
In that case profiling is performed on non-equivalent nuisance manifolds and residual mismodelling can translate into artificial tension, including spurious shifts of the inferred Wilson coefficients, as explicitly illustrated in Section~\ref{sec:sec5:bias}. Note that similar effects may also manifest themselves from the treatment of other shared parameters, such as common systematic uncertainties.

These studies are enabled by \textsc{Redist}, which embeds event-by-event theory reweighting directly into a portable \textsc{pyhf}/\textsc{HistFactory} likelihood. 
Variations of $\theta=(\vec C,\vec\alpha)$ are propagated coherently through acceptance, resolution, and reconstructed template shapes without repeated detector simulation, while the backend-agnostic design permits stringent validation by swapping theory engines (e.g.\ \textsc{EOS}) without altering the statistical model.

Looking forward, the path to sharpening the EFT program is clear. 
Adding a small set of high-leverage discriminants (notably $q^2$- and angular-sensitive information) and refining hadronic-$\tau$ categories (such as separating $\tau\!\to\!\pi\nu$ and $\tau\!\to\!\rho\nu$) will further reduce residual degeneracies. 
Promoting additional shared hadronic inputs, including richer $B\to D^{(*)}$ form-factor shape information and improved $B\to D^{**}\ell\nu$ feed-down modelling, to correlated nuisances will move the framework toward end-to-end experimental realism.
Finally the inclusion of \emph{CP-odd} observables, such as CP asymmetries and signed triple-product correlations, which directly break the approximate ${\rm Im}(C)\!\to\!-{\rm Im}(C)$ ambiguities intrinsic to CP-averaged fits and restore linear sensitivity to imaginary parts of the Wilson coefficients.

In conclusion, portable, morphable likelihoods with explicit shared hadronic nuisance parameters should be regarded as the \emph{default} for future joint LHCb--Belle~II EFT interpretations of $B\to D^{(*)}\tau\nu$. 
They enforce a single consistent nuisance framework, make use of channel complementarity, and eliminate combination artefacts, thereby turning increasing experimental precision into robust constraints on the short-distance physics.

%% file: acknowledgements.tex
This work was funded by the Deutsche Forschungsgemeinschaft (DFG, German Research Foundation) under Germany’s Excellence Strategy – EXC 3107 – Project-ID 533766364. 
We thank Dean Robinson and Michele Papucci for the useful discussions on the HAMMER software.
We thank Frédéric Blanc and Vladimir Gligorov for the careful read of the manuscript.
MP acknowledges support by the German Research Foundation (DFG) Emmy-Noether Grant No. 526218088.
LG acknowledges support by the German Research Foundation (DFG), Grant No. 460248186 (PUNCH4NFDI).
AM acknowledges support by CERN.
BM acknowledges support by UK Science and Technology Facilities Council and the University of Manchester. 
JA, MC and BM acknowledge support by BMFTR under grant no. 05H21PECL1 within ErUM-FSP T04.  

%% file: sec7.tex
In semileptonic decays such as $B^0\!\to D^\ast \ell \nu$ (and in channels with $\tau$ leptons), at least one neutrino escapes detection. To form kinematic observables (e.g.\ $q^2$ or approximate rest-frame variables) we therefore construct a proxy for the parent momentum and infer the missing four-momentum by four-vector subtraction.

The reconstruction strategies used in this work fall into two categories:
\begin{itemize}
  \item an approximate $B$ rest-frame construction for $B^0\!\to D^\ast \ell \nu$ based on the measured $B$ flight direction and a longitudinal-boost ansatz;
  \item a mass-constrained solution for decay chains with a single neutrino at a given stage (e.g.\ $\tau\!\to X_{\rm had}\nu$), which can then be propagated to the $B$ level once the intermediate state is reconstructed.
\end{itemize}

\subsection{Approximate $B$ rest frame from a longitudinal-boost ansatz}
For $B^0\!\to D^\ast \ell \nu$ we define an approximate $B$ momentum by combining the measured $B$ flight direction with a one-dimensional ansatz for the longitudinal boost.

\paragraph{Flight direction.}
From the reconstructed PV and $B$-decay SV positions, the (unit) flight direction is
\begin{equation}
\hat n_B \;=\; \frac{\vec r_{\rm SV}-\vec r_{\rm PV}}{\big|\vec r_{\rm SV}-\vec r_{\rm PV}\big|}
\;\equiv\; (n_x,n_y,n_z)\,.
\end{equation}

\paragraph{Longitudinal boost (``$p_z$ scaling'').}
Let the visible system be $p_{\rm vis}=p_{D^\ast}+p_\ell$ with reconstructed invariant mass $m_{\rm vis}$ and longitudinal momentum $p^{\rm vis}_z$.
We approximate the $B$ longitudinal momentum by assuming the visible system and the $B$ share the same boost along the beam ($z$) axis:
\begin{equation}
p_z^B \;=\; \frac{m_B}{m_{\rm vis}}\,p^{\rm vis}_z\,,
\end{equation}
where $m_B$ is the nominal $B^0$ mass.

\paragraph{Full $B$ momentum and missing four-momentum proxy.}
Imposing that $\vec p_B$ points along $\hat n_B$ fixes the magnitude,
\begin{equation}
|\vec p_B| \;=\; \frac{p_z^B}{n_z}\,,
\qquad\Rightarrow\qquad
\vec p_B \;=\; |\vec p_B|\,\hat n_B\,,
\end{equation}
with $E_B=\sqrt{|\vec p_B|^2+m_B^2}$. The neutrino proxy is then
\begin{equation}
p_\nu^{\rm (proxy)} \;=\; p_B - p_{\rm vis}\,,
\qquad p_B \equiv (E_B,\vec p_B)\,.
\end{equation}

\paragraph{Comments.}
This construction is intentionally simple and stable, but it is not an exact kinematic solution: it neglects event-by-event transverse recoil and therefore induces a non-trivial resolution (and potential bias) on missing-energy observables. In our sensitivity studies this is handled through response modelling in template-based inference.

\subsection{Mass-constrained reconstruction with a quadratic ambiguity}
This method is used when a decay stage contains exactly one invisible particle. In practice, we apply it first to hadronic $\tau$ decays such as $\tau\!\to X_{\rm had}\nu$, using the $\tau$ flight direction from vertexing, and subsequently (once $p_\tau$ is reconstructed) to the parent $B$ decay by treating the reconstructed system $D^\ast+\tau$ as the visible state.

\paragraph{Generic setup.}
Consider a two-body decay of a parent $P$ into a visible system $a$ and a single (massless) neutrino,
\[
P \to a\,\nu,\qquad p_a=(E_a,\vec p_a),\qquad m_a^2=p_a^2,\qquad m_P \text{ known.}
\]
Let the parent flight direction be $\hat n_P$ (from vertexing), and write the unknown parent three-momentum as
\begin{equation}
\vec p_P = p\,\hat n_P,\qquad E_P=\sqrt{p^2+m_P^2}\,,
\end{equation}
with unknown magnitude $p\equiv|\vec p_P|$.
Define the components of $\vec p_a$ parallel/perpendicular to $\hat n_P$,
\begin{equation}
p_{a\parallel} \equiv \vec p_a\cdot \hat n_P,
\qquad
p_{a\perp}^2 \equiv |\vec p_a|^2 - p_{a\parallel}^2.
\end{equation}

\paragraph{Quadratic equation and solutions.}
Imposing a massless neutrino, $(p_P-p_a)^2=0$, yields a quadratic equation for $p$ with two solutions:
\begin{equation}
p \;=\;
\frac{(m_P^2+m_a^2)\,p_{a\parallel}
\;\pm\;
E_a\,\sqrt{(m_P^2-m_a^2)^2 - 4 m_P^2\,p_{a\perp}^2}}
{2\,(E_a^2 - p_{a\parallel}^2)}\,.
\label{eq:quad_solution}
\end{equation}
The discriminant defines the physical region:
\begin{equation}
(m_P^2-m_a^2)^2 - 4 m_P^2\,p_{a\perp}^2 \;\ge\; 0\,.
\end{equation}
Equivalently, in terms of the opening angle $\theta$ between $\vec p_a$ and $\hat n_P$,
\begin{equation}
\sin\theta_{\max} \;=\; \frac{m_P^2 - m_a^2}{2m_P\,|\vec p_a|}\,.
\end{equation}

\paragraph{Resolving the ambiguity in this study.}
In ideal kinematics both solutions in Eq.~\eqref{eq:quad_solution} are admissible; in real data detector resolution can also render the discriminant slightly negative.
To obtain a single-valued proxy that is always defined, we adopt a conservative prescription: we set $\theta=\theta_{\max}$ (i.e.\ clamp the discriminant to zero), so the two branches coincide and
\begin{equation}
p \;\equiv\; |\vec p_P|
\;=\;
\frac{(m_P^2+m_a^2)\,p_{a\parallel}}
{2\,(E_a^2 - p_{a\parallel}^2)}
\bigg|_{\;\theta=\theta_{\max}}\,.
\end{equation}
This removes the discrete ambiguity at the cost of introducing an approximation that is absorbed into the response modelling used in our template-based sensitivity studies.